\documentclass[prd,amsmath,amsfonts,floatfix,nofootinbib,preprintnumbers,superscriptaddress,showpacs]{revtex4}
\usepackage{amsfonts}
\usepackage{graphicx}
\usepackage{amsmath}
\usepackage[usenames]{color}
\usepackage{slashed}
\usepackage[caption=false]{subfig}	

\graphicspath{{figures/StrongYukawaCoupling/}{figures/finiteT_2/}{figures/}}
\usepackage{xcolor}
\usepackage{float}

\newcommand*{\chpt}{\raise0.4ex\hbox{$\chi$}PT}

\newcommand{\Delstar}{\ensuremath{\Delta^{\raise0.18ex\hbox{${\scriptstyle *}$}}}}
\def\gtwid{{\,\raise.35ex\hbox{$>$\kern-.75em\lower1ex\hbox{$\sim$}}\,}}
\def\ltwid{{\,\raise.35ex\hbox{$<$\kern-.75em\lower1ex\hbox{$\sim$}}\,}}
\def\leftvec{{\raise1.5ex\hbox{$\leftarrow$}\kern-1.00em}}
\def\rightvec{{\raise1.5ex\hbox{$\rightarrow$}\kern-1.00em}}
\def\half{{\scriptstyle \raise.2ex\hbox{${1\over2}$}}}
\def\threehalves{{\scriptstyle \raise.15ex\hbox{${3\over2}$}}}
\def\third{{\scriptstyle \raise.15ex\hbox{${1\over3}$}}}
\def\third{{\scriptstyle \raise.15ex\hbox{${1\over3}$}}}
\def\twothirds{{\scriptstyle \raise.15ex\hbox{${2\over3}$}}}
\def\fourth{{\scriptstyle \raise.15ex\hbox{${1\over4}$}}}

\def\op{{\mathcal{O}}}

\def\beq{\begin{equation}}
\def\eeq{\end{equation}}
\def\barr{\begin{array}}
\def\earr{\end{array}}

\newcommand*{\bea}{\begin{eqnarray}}
\newcommand*{\eea}{\end{eqnarray}}
\newcommand*{\be}{\begin{equation}}
\newcommand*{\ee}{\end{equation}}

\begin{document}

\preprint{ \vbox{\hbox{CERN-PH-TH/2012-255}} }
\preprint{ \vbox{\hbox{DESY 12-162}} }
\begin{titlepage}
\begin{flushright}
\end{flushright}
\vskip 0.5cm
\begin{center}
{\Large \bf Higgs-Yukawa model in chirally-invariant lattice field theory\\} 
\vskip1cm {\large\bf John~Bulava$^{a}$, Philipp~Gerhold$^{b,c}$, 
Karl~Jansen$^{c}$, Jim~Kallarackal$^{b,c}$, Bastian~Knippschild$^{d}$,
C.-J. David~Lin$^{e,f}$, Kei-Ichi~Nagai$^{g}$, Attila~Nagy$^{b,c}$, Kenji~Ogawa$^{h}$}\\ \vspace{.5cm}
{\normalsize {\sl 
$^{a}$ CERN, Physics Department, 1211 Geneva 23, Switzerland\\
$^{b}$ Instit\"{u}t f\"{u}r Physik, Humboldt-Universit\"{a}t zu
Berlin, D-12489 Berlin, Germany\\
$^{c}$ NIC, DESY, Platanenallee 6, Zeuthen D-15738, Germany\\
$^{d}$ Department of Physics, National Taiwan University, Taipei 10617, Taiwan\\
$^{e}$ Institute of Physics, National Chiao-Tung University, Hsinchu 300, Taiwan\\
$^{f}$ Division of Physics, National Centre for Theoretical Sciences, Hsinchu 300, Taiwan\\
$^{g}$ Kobayashi-Maskawa Institute, Nagoya University, Nagoya, Aichi 464-8602, Japan\\
$^{h}$ Department of Physics, Chung-Yuan Christian University,
Chung-Li 32023, Taiwan\\ 
}}

\vskip1.0cm {\large\bf Abstract\\[10pt]} \parbox[t]{\textwidth}{{
Non-perturbative numerical lattice studies of the Higgs-Yukawa sector of the standard model 
with exact chiral symmetry are reviewed.  In particular, we discuss bounds on the 
Higgs boson mass at the standard model top quark mass, and
in the presence of heavy fermions. We present a comprehensive study of the phase 
structure of the theory at weak and very strong values of the Yukawa coupling as well 
as at non-zero temperature. 
}}
\end{center}
\vskip0.5cm
{\small PACS numbers: 12.38.Gc, 12.39.Fe, 12.39.Hg, 14.20.Mr, 14.40.Nd}
\end{titlepage}

\pacs{11.15.Ha,11.30.Rd,12.38.Gc,12.39.Fe,12.39.Hg}


\section{Introduction}\label{sec:intro}
The Higgs-Yukawa sector of the Standard Model (SM) describes the generation
of fermion masses via the non-vanishing vacuum expectation value
($vev$) acquired by the Higgs field
which couples 
through a Yukawa coupling to the fermions. 
The essential element in this picture is that the coupling of 
the fermions to the Higgs field is chirally 
invariant which leads to the gauge invariant electroweak 
sector of the SM   
in the presence of gauge fields. 

There are two couplings in the Higgs-Yukawa sector.  They are associated with
the Yukawa and the quartic scalar self-interaction operators.
These couplings are directly related to the fermion and the 
Higgs boson masses, respectively. In the scenario that these masses
are large, the corresponding couplings grow strong, and it becomes unclear whether the theory can be 
analysed using perturbation theory or whether
non-perturbative methods must be employed. 
There are indeed examples  
where the applicability of perturbation theory is  
questionable. The first is the upper Higgs boson mass bound which is 
based on 
triviality arguments~\cite{Dashen:1983ts}. Here the Higgs boson mass can become 
large, resulting in a strong value of the quartic coupling such 
that perturbation theory may not work anymore. The second is 
the lower Higgs boson mass bound which is based on vacuum instability 
arguments~\cite{Lee:1974ma,Linde:1975sw,Weinberg:1976pe,Sher:1988mj}. 
Here it is unclear whether this instability is not an 
artefact of perturbation theory applied at large values of the 
Higgs field such that an expansion around the minimum of the effective 
potential is not justified anymore. 

It is important to stress that both the lower and the upper Higgs boson 
mass bounds are intrinsically related to the cut-off of the theory. 
Thus, a calculation of the Higgs boson mass bounds can in turn 
be used to determine the cut-off up to which the SM 
is valid, once the SM Higgs boson mass has been determined. 
If, for example, the recent result for a scalar particle 
at the Large Hadron Collider (LHC)~\cite{ATLAS,CMS} 
is confirmed as a 
SM Higgs boson with a mass of about $125\text{ GeV}$, the SM
could be valid up to very high energies before violating 
the Higgs boson mass bounds, see Ref.~\cite{Degrassi:2012ry} for a recent 
analysis at next-to-next leading order of perturbation theory. 

Another example where non-perturbative calculations are necessary is 
the possibility of a heavy fourth fermion generation~\cite{Carena:2004ha,Holdom:2009rf}
which would lead 
to a large value of the corresponding Yukawa coupling. 
Besides these concrete examples, 
it is conceptually
very important to study the Higgs-Yukawa sector in a non-perturbative manner 
since
questions such as the phase structure of the model, or the spontaneous
breaking of the $SU(2)\otimes SU(2)$ symmetry which underlies the Higgs mechanism 
are of intrinsically non-perturbative nature. 

The need for a non-perturbative investigation
of the Higgs-Yukawa sector of the SM has been 
realised already in the early 1990's. A natural choice 
of a non-perturbative tool is, of course, Euclidean 
lattice field theory. However, in these early studies, 
the lattice formulations of the Higgs-Yukawa sector were 
lacking a chirally symmetric form of the Yukawa coupling 
term. 
The absence of a chirally invariant Yukawa coupling term in the Lagrangian 
led to severe difficulties in studying Higgs-Yukawa model 
on the lattice,  
see Refs.~\cite{Smit:1989tz,Shigemitsu:1991tc,Golterman:1990nx,De:1991me,
montvay1997quantum,Golterman:1992ye,Jansen:1994ym}
and references therein.                                      

The situation changed however, when it was realised that 
--based on the Ginsparg-Wilson relation~\cite{Ginsparg:1981bj}--
there exists
a consistent formulation of an exact lattice chiral 
symmetry~\cite{Luscher:1998pqa},
which allows the chiral
character of the Higgs-fermion coupling structure of the SM to be preserved on the lattice
in a conceptually fully controlled manner.
This triggered a number of lattice investigations of Higgs-Yukawa like
models~\cite{Bhattacharya:2006dc,Giedt:2007qg,Poppitz:2007tu,Gerhold:2007yb,Gerhold:2007gx,Fodor:2007fn,Gerhold:2009ub,Gerhold:2010wy,Gerhold:2010bh}.

In this article, we report on the status of the lattice Higgs-Yukawa model
using a lattice formulation that obeys an  
exact lattice chiral symmetry as will be explained in Sec.~\ref{sec:lattice_setting}. 
In Sec.~\ref{sec:higgs_mass_bounds} we will provide results 
for the lower and upper Higgs boson mass bounds as well as the 
resonance parameters
of the Higgs boson~\cite{Gerhold:2009ub,Gerhold:2010bh,Gerhold:2010wy,Gerhold:2011mx}. 
We also extend the study of the Higgs boson mass bounds to the 
case of a fourth quark generation~\cite{Gerhold:2010wv}.
This calculation will result in rather severe constraints on the existence of 
a fourth fermion generation. 

This article is organised as the following.  In
Sec.~\ref{sec:lattice_setting}, we describe the setting of our lattice
simulations.  Section~\ref{sec:higgs_mass_bounds} contains results of
our work on the Higgs boson mass bounds in the Higgs-Yukawa model.  In
particular, we have investigated the effects of the fermion mass on
these bounds.  In Sec.~\ref{sec:phase_structure_results}, we present our study of the
phase structure of the model.  These include the bulk phase
transitions at small values of the bare Yukawa
coupling~\cite{Gerhold:2007yb,Gerhold:2007gx}, as well as in the
regime of strong-Yukawa coupling~\cite{Bulava:2011jp}.  We also show
results and the status of our work on the finite-temperature phase
structure in Sec.~\ref{sec:finite_t}.  Finally, we conclude in
Sec.~\ref{sec:summary}.

All statistical errors we quote in this article
were obtained with a jackknife or bootstrap analysis, taking
possible effects of autocorrelations fully into account.  Statistical errors 
of the results presented in Sec.~\ref{sec:bulk_phase_transition} have
also been cross-checked using the method in Ref.~\cite{Wolff:2003sm}.

%

\section{Lattice setting and simulation strategy}\label{sec:lattice_setting}
\subsection{The action}
\label{sec:the_action}
The Euclidean action of the continuum Higgs-Yukawa model containing one doublet of fermions,
denoted as $t^{(c)}$ and $b^{(c)}$,
and a complex scalar doublet, $\varphi^{(c)}$,  is
\begin{eqnarray}
\label{eq:action_continuum}
&&S^{\text{cont}}[\bar{\psi^{(c)}}, \psi^{(c)}, \varphi^{(c)}] = \int d^4 x \left\{\frac{1}{2}\left(\partial_{\mu} \varphi^{(c)} \right)^{\dagger} 
	\left(\partial^{\mu} \varphi^{(c)} \right) 
	+  \frac{1}{2} m_0^2 \varphi^{(c)\dagger} \varphi^{(c)} + \frac{\lambda_{0}}{4} \left(\varphi^{(c)\dagger} \varphi^{(c)} \right)^2 \right\} \nonumber\\
&&	\hspace{3.5cm} +\int d^4 x  \left\{\overline{t^{(c)}} \slashed \partial t^{(c)} + \overline{b^{(c)}} \slashed \partial b^{(c)} +
			y_{b_0}\overline{\psi^{(c)}_{L}} \varphi^{(c)}\, {b^{(c)}_{R}} +
			y_{t_0}\overline{\psi^{(c)}_{L}} \tilde \varphi^{(c)}\, {t^{(c)}_{R}}
			+ h.c. \right\} ,  \\
&& \mathrm{where}\mbox{ }  \tilde{\varphi}^{(c)}=i\tau_2 \varphi^{(c)}
\mbox{ }(\tau_{i} \mbox{ }{\rm are}\mbox{ }{\rm the} \mbox{ }{\rm
  Pauli} \mbox{ }{\rm matrices}),
\nonumber \\
&& \hspace{1cm} \psi^{(c)}_{L} = P_{-} \psi^{(c)} = P_{-} \left
  ( \begin{array}{c} t^{(c)} \\ b^{(c)} \end{array} \right ) =
\left ( \frac{1-\gamma_{5}}{2} \right )\left ( \begin{array}{c} t^{(c)} \\
    b^{(c)} \end{array} \right ) , \nonumber \\
&& \hspace{1cm} t_{R}^{(c)} = P_{+} t^{(c)}  = \left ( \frac{1 +
    \gamma_{5}}{2} \right ) t^{(c)} , \mbox{ } {\mathrm{and}} 
\mbox{ }{\mathrm{similar}}\mbox{ }{\mathrm{for}}\mbox{ } b^{(c)}_{R}
. \nonumber
\end{eqnarray}
In the above equation, $m_0$ is the bare mass, $\lambda_0$ labels the
bare quartic coupling, and $y_{t_0/b_0}$ denote the bare Yukawa
couplings.  The superscript, $(c)$, in the scalar and spinor fields indicates
that these are dimensionful variables defined in the continuum.
Here we stress that gauge fields are not included in our study, and we
perform calculations for only one doublet of fermions throughout
this work.
Moreover, if not stated otherwise, the Yukawa couplings $y_{t_0}$ and $y_{b_0}$
are set equal.
It is straightforward to discretise the pure-scalar component of the
above action to obtain
\begin{equation}
\label{eq:scalar_action_lattice}
 S_{\Phi}^{\rm latt} =  \sum_{\alpha=1}^{4} \left \{ - \sum_{x,\mu}
   \Phi_{x}^{\alpha} \Phi_{x+\hat{\mu}}^{\alpha}  + \sum_{x} \left [
   \frac{1}{2} (8+\bar{m}^{2}_{0}) \Phi^{\alpha}_{x} \Phi_{x}^{\alpha}  +
   \frac{\lambda_{0}}{4} \left ( \Phi^{\alpha}_{x} \Phi_{x}^{\alpha}
   \right )^{2} \right ]\right \} , 
\end{equation}
where $x$ is a site on the space-time lattice.  
The symbol $\hat{\mu}$ denotes the unit vector in the
space-time direction $\mu$. 
The mass parameter, $\bar{m}_{0} = a m_{0}$ with $a$ being the lattice spacing, is
dimensionless.  The real-valued field variables,
$\{ \Phi_{x}^{\alpha} \}$, are rendered dimensionless by a proper
rescaling with $a$, and are defined on all lattice sites.
These field variables are related to the discretised version of the
complex scalar doublet, $\varphi^{(c)}$, in
Eq.~(\ref{eq:action_continuum}) through
\begin{equation}
 a \varphi^{({\rm latt})} = \left ( \begin{array}{c} \Phi^{2} + i
     \Phi^{1}  \\ \Phi^{4} - i \Phi^{3}  \end{array} \right )  .
\end{equation}

It is convenient to rewrite the scalar action in
Eq.~(\ref{eq:scalar_action_lattice}) as
\begin{equation}
\label{eq:action_lattice_bosonic}
	S_{\phi}^{\rm latt}  = \sum_{\alpha = 1}^{4} \left \{  - 2 \kappa
          \sum_{x,\mu}  \phi_{x}^{\alpha}
          \phi_{x+\hat{\mu}}^{\alpha} + 
				\sum_{x}   \left [ \phi_{x}^{\alpha}
                                \phi_{x}^{\alpha} + \hat{\lambda}
                                \left ( \phi_{x}^{\alpha}
                                  \phi_{x}^{\alpha} - 1\right )^{2}
                              \right ] \right \} ,
\end{equation}
with the change of variables,
\begin{equation}
	 \Phi^{\alpha} = \sqrt{2 \kappa} \phi^{\alpha},\quad
	 \lambda_0 = \frac{\hat{\lambda}}{{\kappa^2}},\quad
	 \bar{m}_0^2 = \frac{1 - 2 \hat{\lambda} -8 \kappa}{\kappa}.
\end{equation}
For the fermions we use the action
\begin{equation}\label{eq:action_lattice_fermionic}
	S_{f}^{\rm latt} = \sum\limits_{x} \bar{\psi}_x\left[ D^{ov} +
          P_{_+} \phi^{\alpha}_{x} \theta_{\alpha}^{\dagger} \text{diag}(\hat{y}_t,\hat{y}_b) 
	\hat{P}_{_+} + 
						P_{_-}
                                                \text{diag}(\hat{y}_t,\hat{y}_b)
                                                \phi^{\alpha}_{x} \theta_{\alpha}
                                                \hat{P}_{_-} \right]
                                              \psi_x ,
\end{equation}
where $\hat{y}_{t/b} = \sqrt{2 \kappa} y_{t_0/b_0}$, and 
\beq
 \theta_{1,2,3} = -i \tau_{1,2,3}, \mbox{ } \theta_{4} = 1_{2\times 2}\; ,
\eeq
where a summation over $\alpha$ is understood.
The dimensionless
spinor field $\psi$ is,
\begin{equation}
 \psi = a^{\frac{3}{2}} \left ( \begin{array}{c} t^{({\rm latt})} \\ b^{({\rm latt})} \end{array} \right ) ,
\end{equation}
with $ t^{({\rm latt})}$ and $b^{({\rm latt})}$ being the lattice
version of $t^{(c)}$ and $b^{(c)}$. 
For the fermion kinetic term, we use the 
overlap operator
\cite{Neuberger:1997fp, Neuberger:1998wv, Hernandez:1998et},
\begin{equation} \label{eq:DefOfNeuberDiracOp}
 D^{ov}  =  \rho \left\{1+\frac{ A}{\sqrt{ A^\dagger  A}}   \right\},  \quad A = D^{W} - \rho, 
\end{equation}
where $\rho$ is a free, dimensionless parameter, restricted to $0 <
\rho < 2r$.  
The locality properties of the overlap operator are optimal for
$\rho=1$ in the case of vanishing gauge
couplings~\cite{Hernandez:1998et}, and therefore we set it to this
value in this work.
The operator $D^{W}$ denotes the Wilson Dirac 
operator defined as 
\begin{equation}
\label{eq:DefOfWilsonOperator}
D^{W} = \sum\limits_\mu \gamma_\mu \nabla^s_\mu - \frac{r}{2} \nabla^b_\mu\nabla^f_\mu,
\end{equation}
where $\nabla^{f,b,s}_\mu$ are the (respectively) forward, 
backward and symmetrised lattice nearest-neighbour difference operators in 
direction $\mu$, and the Wilson parameter $r$ is chosen to be $r=1$.
The modified chiral projectors are given by:
\begin{equation}
 \hat{P}_{\pm} = \frac {1 \pm \hat{\gamma}^5}{2}, \quad
 \hat{\gamma}^5 = \gamma^5 \left( 1 - \frac{1}{\rho} D^{ov} \right) .
\end{equation}
This action now obeys an exact global $\mbox{SU}(2)_L\times
\mbox{U}(1)_Y$ (with $Y$ being the hyper-charge)
lattice chiral symmetry with the transformations:
\begin{equation}
 \psi \rightarrow  U_Y \hat P_+ \psi + U_Y\Omega_L \hat P_- \psi ,\quad
 \bar\psi \rightarrow \bar\psi P_+ \Omega_L^\dagger U_{Y}^\dagger + \bar\psi P_- U^\dagger_{Y}, \quad
\phi \rightarrow U_Y  \phi \Omega_L^\dagger, \quad
\phi^\dagger \rightarrow \Omega_L \phi^\dagger U_Y^\dagger,
\end{equation}
for any $\Omega_L \in \text{SU}(2)_L$ and $U_Y \in \text{U}(1)_Y$.
%


\subsection{Implementation}
\begin{figure}[H]
	\begin{center}
	$	
	\begin{array}{ccc}
		\hspace*{-0.4cm} \includegraphics[width=0.33\linewidth]{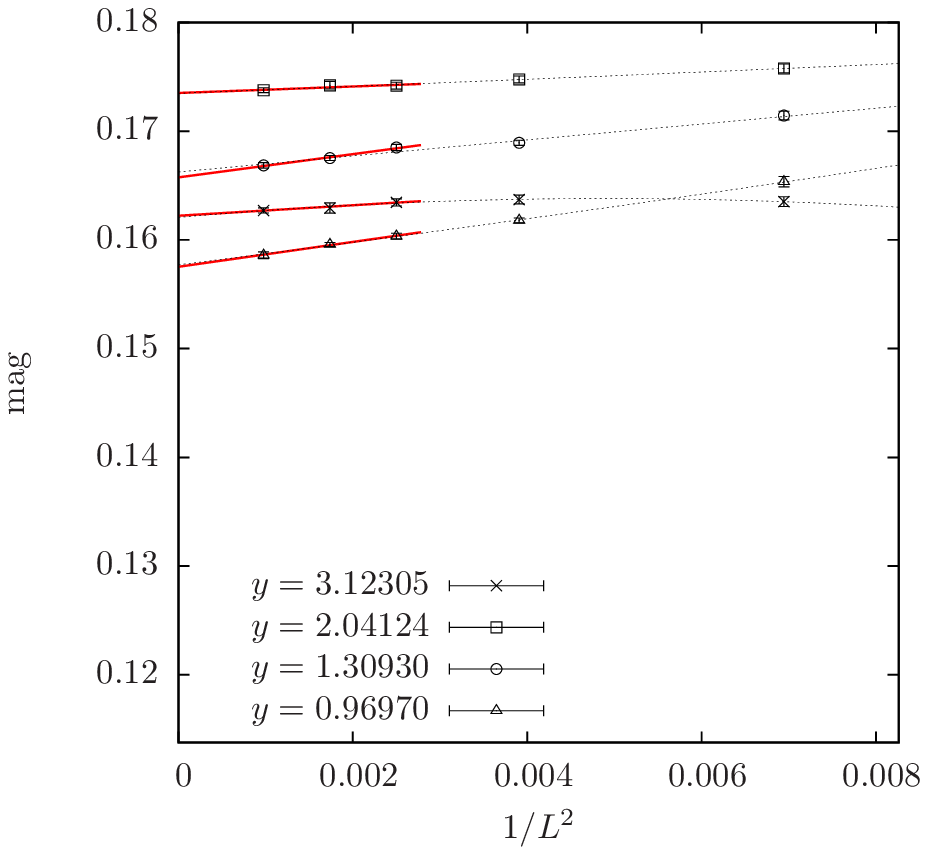}  &
		\hspace*{-0.3cm}  \includegraphics[width=0.33\linewidth]{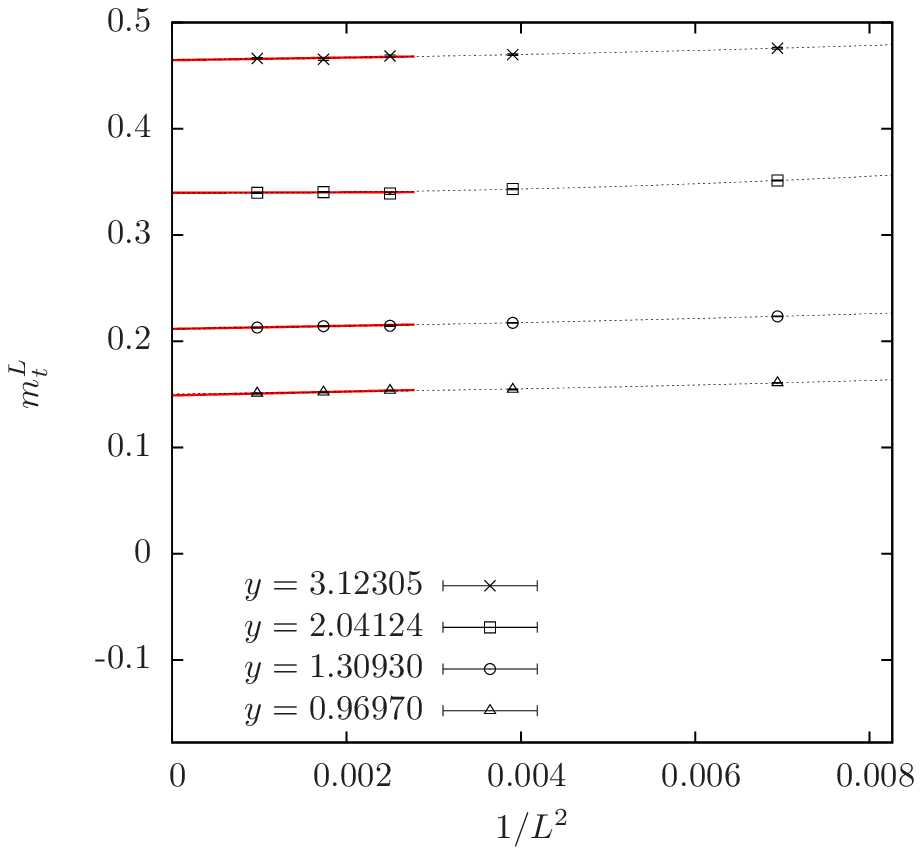} &
		\hspace*{-0.3cm}  \includegraphics[width=0.33\linewidth]{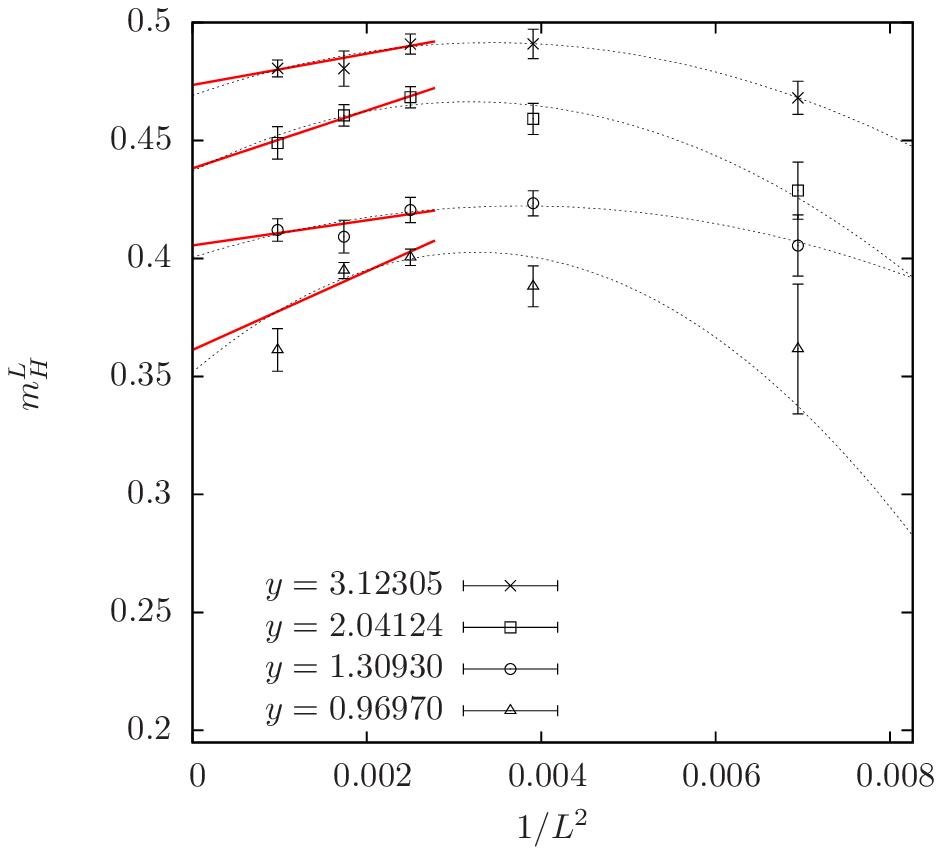} \\
	\end{array}
	$	
	\end{center}
\vspace*{-0.5cm}
\caption{Finite volume effects in the
  magnetisation as defined in Eqs.~(\ref{eq:bare_vev_definition}) and (\ref{eq:mag})
  (left), the fermion mass (middle), and the Higgs boson mass (right) at a cut-off around $1.5\mathrm{TeV}$. The data 
are obtained at infinite bare scalar-quartic coupling, $\hat{\lambda}$, 
and fermion masses in the range $m_{f} \approx 200 - 700\mathrm{GeV}$. 
The lattice sizes used are $L=L_s=12,16,20,24,32$. We show linear (solid lines) and 
quadratic (dotted lines) fits in $1/L^2$.}
\label{fig:finiteVolumeExamples}
\end{figure}

The actions in Eqs.~(\ref{eq:action_lattice_bosonic}) and
(\ref{eq:action_lattice_fermionic}) are used in our numerical
simulations.
We perform calculations on
asymmetric 4-dimensional lattice volumes
\beq
 \label{eq:4V_def}
 V_{4} = L_{s}^{3} \times L_{t} ,
\eeq
where $L_{s}$ and $L_{t}$ are dimensionless spatial and temporal
lattice sizes, respectively.  In all our zero-temperature computations, we choose
\beq
 \label{eq:asym_vol}
 L_{t} = 2 L_{s} = 2 L ,
\eeq
with $L$ typically ranging from 8 to 32.  We stress that it is
essential to perform computations for the Higgs-Yukawa models on large
volumes.  This is because
the Goldstone bosons are (almost) massless and induce
significant finite-size effects proportional to $L^{-2}$, in contrast to the 
exponential effects known for a single-particle spectrum and matrix
elements for theories such as QCD with massive quarks.
Figure~\ref{fig:finiteVolumeExamples} shows some examples of finite-volume
effects that are present in quantities investigated in this work.
It is clear from these plots that finite volume effects can be very
large in the calculation of the Higgs boson mass, while they may be
mild in other quantities.

We implement the polynomial Hybrid
Monte Carlo (pHMC) algorithm~\cite{Frezzotti:1997ym, Frezzotti:1998eu, Frezzotti:1998yp}, with various improvements 
(see Ref.~\cite{Gerhold:2009zf} for a summary), to perform
non-perturbative calculations of the path integral. When compared 
to simulations in QCD using overlap fermions \cite{Hashimoto:2008fc}, 
it is the absence of gauge fields 
that makes the application of the overlap operator numerically 
feasible even on large lattices, as it is diagonal in momentum space.

\subsection{Basic observables}
\label{sec:basic_observables}
As described in Sec.~\ref{sec:the_action}, our simulations are
performed using only dimensionless variables in the action.  This is
achieved by rescaling all the dimensionful quantities with appropriate
powers of the lattice spacing, $a$.   Therefore, to make connection to
the real world and to have basic understanding of the spectrum of the
theory, it is essential to determine the lattice spacing.
This is normally carried out by computing the $vev$ of the scalar
field, and then setting it to the value of 246GeV.  Before we describe
the details of this procedure, it should be noticed that the scalar 
$vev$ is always zero in a finite system.
In principle, one would have to introduce an external source that couples to the 
scalar field and breaks the O(4) symmetry explicitly, and perform the 
infinite-volume extrapolation for every quantity computed on the lattice,
before taking the source to zero.   However, this procedure is
numerically very demanding, and we resort to an alternative method in
which we ``rotate'' the complex scalar doublet in every field
configuration, such that its ensemble average is given by
\begin{equation}
\label{eq:bare_vev_definition}
 \langle \hat{\phi}_{\text{rot}} \rangle = \left( \begin{array}{c} 0 \\ v \end{array}
 \right), \quad v=\sqrt{2\kappa} \langle m \rangle ,
\end{equation}
with 
\beq
\label{eq:mag}
m = \frac{1}{V_{4}} \sum_{x} \left ( \sum_{\alpha}
  |\phi_{x}^{\alpha}|^{2} \right )^{1/2} 
\eeq
defined on each configuration.  It can be shown that the
magnetisation, $\langle m \rangle$, is equivalent to the scalar $vev$
in the infinite-volume limit~\cite{Hasenfratz:1989ux, Hasenfratz:1990fu, Gockeler:1991ty}.

The renormalised scalar $vev$ is given by
\begin{equation}
 v_r=\frac{v}{\sqrt{Z_G}},
\end{equation}
where $Z_G$ is the Goldstone-boson wavefunction renormalisation
constant.  This renormalisation constant, and the Higgs-field wavefunction
renormalisation constant $Z_H$, can be 
extracted from the momentum-space Euclidean propagators of the corresponding
bosons \cite{Gerhold:2009ub,Gerhold:2010bh},
\bea
 && G_{G/H}(p^{2}) = \frac{1}{L_{t}^{2}\cdot L^{6}_{s}} \sum_{t_{x}, t_{y}}
 \sum_{\vec{x},\vec{y}}  {\mathrm{e}}^{i \vec{p} \cdot (\vec{x} -
   \vec{y}) + i p_{4} (t_{x} - t_{y}) }\mbox{ }
 \left \langle \op_{G/H}(\vec{x},t_{x}) \mbox{ }
   \op^{\dagger}_{G/H}(\vec{y},t_{y})\right \rangle \nonumber \\
 && \nonumber\\
\label{eq:GH_mom_prop}
 && \hspace{1.5cm} \stackrel{p^2 \ll 1}{\longrightarrow}
 \frac{Z_{G/H}}{p^{2} + m^{2}_{G/H}} ,
\eea
with $\op_{G/H}$ being the Goldstone and Higgs fields, respectively,
and all the masses and momenta are in lattice units.  

Through the investigation of the momentum dependence of the Goldstone
boson propagator,  $Z_{G}$ can be determined.    This procedure can be
improved by performing calculations in one-loop lattice perturbation
theory and obtaining the propagators to this order\cite{Jim:thesis}.
The lattice spacing, which is related to the inverse of the cut-off
scale, $\Lambda$, can now be obtained in natural units with,
\begin{equation}
\label{eq:setting_a}
 a = \Lambda^{-1},\quad \Lambda=\frac{246 \text{GeV}}{v_r} .
\end{equation}

The masses of the bosons are given by the pole of the
Euclidean propagators in Eq.~(\ref{eq:GH_mom_prop}).  They can also be 
extracted from the time dependence of the Euclidean correlators with
zero spatial momentum \cite{Gerhold:2009ub,Gerhold:2010bh},
\bea
 && C_{G/H} (\Delta t) = \frac{1}{L_{t}\cdot L_{s}^{6}} \sum_{t}
 \sum_{\vec{x},\vec{y}} \left \langle
 \op_{G/H} (\vec{x}, t + \Delta t) \mbox{ }\op^{\dagger}_{G/H}(\vec{y}, t)
\right \rangle \nonumber\\
 && \nonumber\\
 && \hspace{1.5cm} \stackrel{\Delta t \gg 1}{\longrightarrow}
 A_{G/H}\mbox{ }
 {\mathrm{exp}} \left ( \frac{-m_{G/H} L_{t}}{2} \right )\mbox{ }
 {\mathrm{cosh}}\left [ m_{G/H} \mbox{ } \left ( \frac{L_{t}}{2} -
     \Delta t
    \right  )\right ] \nonumber\\
&&\nonumber\\
\label{eq:GH_corr}
\eea
where $A_{G/H}$ are constants that are proportional to $Z_{G/H}$.
This formula is valid when periodic boundary conditions are imposed.
Here we stress that this method is applicable only when the ground
state is the target single-particle state.   Therefore, one has to be
cautious when studying the Higgs boson, since it may decay into 
even number of Goldstone bosons.  
The unstable nature of the Higgs boson and the calculation of its
resonance parameters 
will be discussed in more detail in
Sec.~\ref{sec:Higgs_mass}.

Finally, to compute the masses of the fermions, we resort to 
the correlator \cite{Gerhold:2009ub,Gerhold:2010bh}
\begin{equation}
 C_f(\Delta t) = \frac{1}{L_t \cdot L_s^6} \sum_{t} \sum_{\vec{x}, \vec{y}} 
                \left \langle  \operatorname{Tr} \left \{ 
					 \hat{P}_{-}\psi (t+ \Delta t, \vec x) \cdot \bar{\psi}(t, \vec y)  P_{-} 
					  \right\} \right \rangle,
\end{equation}
where the trace is over the spinor indices.  By studying the time
dependence of this correlator, 
\begin{equation}
 C_f(\Delta t \gg 1) \propto  {\mathrm{exp}} \left ( \frac{-m_{f} L_{t}}{2} \right )\mbox{ }
 {\mathrm{cosh}}\left [ m_{f} \mbox{ } \left ( \frac{L_{t}}{2} -
     \Delta t
    \right  )\right ] ,
\end{equation}
the fermion mass can be extracted.

\section{Bounds on the Higgs mass}\label{sec:higgs_mass_bounds}
The lattice techniques described in the last section can be applied to 
the calculation of Higgs boson mass 
bounds~\cite{Gerhold:2010bh,Gerhold:2010wv}. In what follows, we study the model in the broken phase, i.e. 
where the $vev$ of the scalar field is non-zero. The Higgs boson mass is 
bounded 
from above by the triviality argument, which reflects the Gaussian nature of the 
fixed point 
of the theory. This bound is not universal and depends logarithmically on the 
UV cut-off of the theory. Indeed variations in the triviality 
bound between 
different lattice regularisations have been observed in the pure $\phi^4$ 
theory~\cite{Heller:1993yv}. 

There is also an argument from perturbation theory 
that the Higgs boson mass is bounded from below by 
a vacuum-stability requirement. 
The picture for the lower bound in perturbation theory arises 
by examining the 
effective potential. As the 
fermion fields contribute negatively to the effective potential, 
they have a destabilising effect. By demanding the stability of the 
theory, this leads then to lower Higgs boson mass bounds. 
However, it is known that the 
perturbative expansion breaks down for Yukawa couplings near or less than  
the tree level unitarity bound~\cite{Denner:2011vt}, which is roughly 
500 to 600GeV~\cite{Chanowitz:1978mv, Chanowitz:1978uj}. 
In addition, the perturbative 
instability occurs at large values of the scalar field where an expansion 
around the minimum of the effective potential may not 
be trustworthy. Therefore it is desirable to have a 
non-perturbative calculation. 

Although also the lower Higgs boson bound is non-universal, 
it is expected that it shows a much milder dependence 
on effects of the regularisation employed since a typical 
ratio $\Lambda/m_H$ is of $O(10)$ for the lower bound, 
while $\Lambda/m_H \sim 0.5$ for the upper bound. 
In the light of the 
recent discovery of a scalar particle at the LHC, the lower bound becomes 
very interesting: if this scalar particle will turn out to be the Higgs boson, 
the lower mass bound can be used to estimate the breakdown scale of the SM,
i.e. the scale where new physics must enter to preserve the stability of 
the theory.

In this work, we compute the upper and lower bounds of the Higgs
boson mass from non-perturbative, direct calculations using lattice
field theory without relying on assumptions such as triviality 
or vacuum instability.
From the study of the pure $\phi^4$ theory, it is 
known~\cite{Hasenfratz:1987eh,Kuti:1987nr,Luscher:1988uq,Hasenfratz:1988kr} 
that the Higgs 
boson mass is 
a monotonically increasing function of the quartic coupling $\lambda$ at 
fixed lattice spacing. This feature has been demonstrated to be
present also in the Higgs-Yukawa 
theory~\cite{Gerhold:2010wy} at fixed value of $m_f$. 
Therefore, in this work the lower bounds for particular values of $m_f$ and $\Lambda$ are
determined at $\hat{\lambda}=0$, while the upper bounds are obtained at $\hat{\lambda}=\infty$.

\subsection{Calculating the Higgs boson mass}
\label{sec:Higgs_mass}
As pointed out in Sec.~\ref{sec:basic_observables}, 
calculating the mass of the Higgs boson is challenging because of its unstable nature, as it 
decays into even numbers of Goldstone bosons.  Extracting the masses
and the widths of 
unstable states in lattice field theory is subtle, because the theory
is formulated in Euclidean space.   It is further complicated by the 
quantisation of spatial momenta in finite volume, since the kinematics
may prevent a resonance state from decaying. Therefore, 
a state which is unstable in infinite volume can remain a stable
eigenstate in finite volume.  

However, below the inelastic threshold, the infinite-volume 
phase shift of two-particle scattering can be determined via the
investigation of finite-size effects in the energy
spectrum~\cite{Luscher:1990ux}. Such finite-volume techniques for
studying scattering states,
albeit very challenging to implement in practice, 
can be used to extract resonance masses and
widths in Euclidean quantum field theory~\cite{Luscher:1991cf}.

\begin{figure}[H]
\begin{center}
\includegraphics[width=0.42\textwidth]{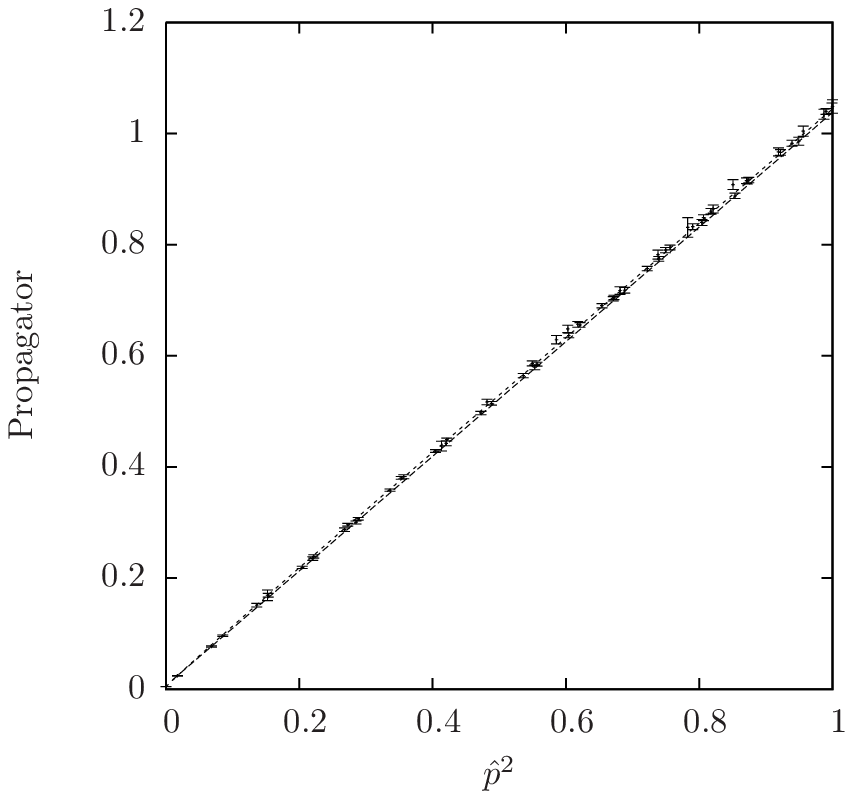}
\includegraphics[width=0.52\textwidth]{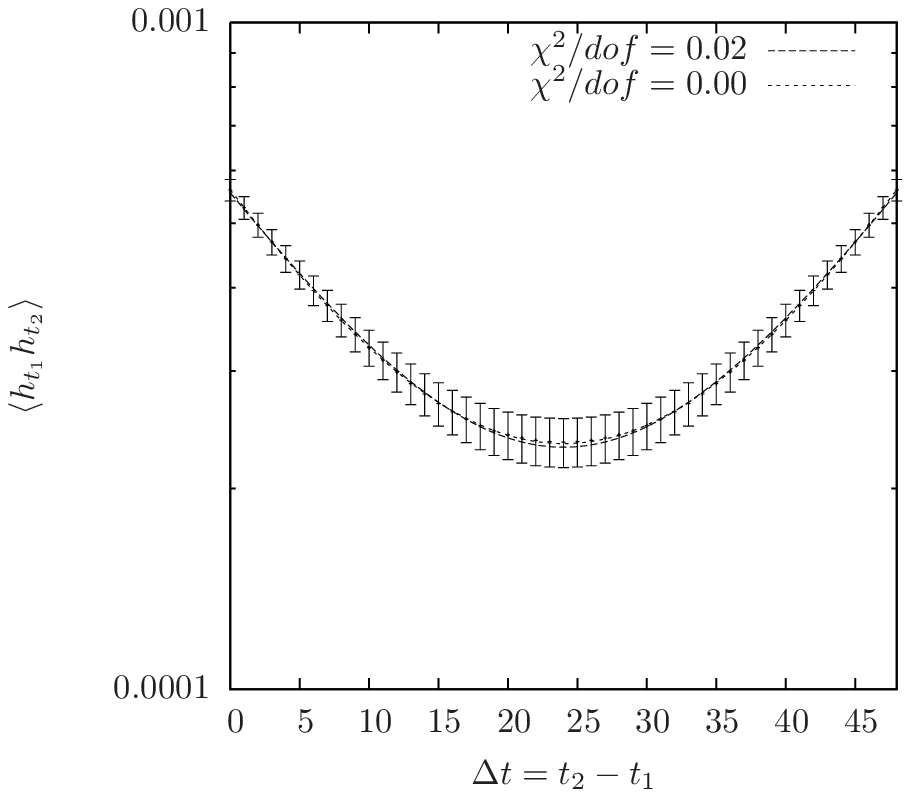}
\end{center}
\caption{Examples of fits to the Higgs momentum space propagator and the 
Higgs temporal correlation function to obtain $m_H^p$ and $m_H^c$, 
respectively. The results are from a $24^{3} \times 48$ lattice with 
$m_{f}=195\mathrm{GeV}$, $\Lambda=1.5\mathrm{TeV}$. The fitted values are 
$m_H^{p} = 96.0(4.3)\mathrm{GeV}$ and 
$m_{H}^c = 96.4(6.9)\mathrm{GeV}$ where the errors are statistical 
only and do not reflect the uncertainty in the scale determination which, however, 
affects both values in the same way.}
\label{fig:fits}
\end{figure}

In this work, we first compute the mass of the Higgs boson by assuming
that its width is zero, therefore it is a stable particle
in finite volume.  To check this assumption, we will later use the
above-mentioned finite-volume method to obtain results of
the Higgs boson width, and confirm that the widths is in fact small thus 
not affecting the results assuming a stable Higgs boson. 
Under the zero width assumption,  
we extract the Higgs boson mass using the two approaches described in 
Sec.~\ref{sec:basic_observables}.  Namely, we study the propagator in
Eq.~(\ref{eq:GH_mom_prop}), and the correlator in
Eq.~(\ref{eq:GH_corr}).  
We then extract the Higgs boson mass by a fit of the propagator to a perturbation 
theory inspired formula \cite{Gerhold:2009ub,Gerhold:2010bh} and by a fit 
to an exponential form of the correlator of Eq.~(\ref{eq:GH_corr}).
The Higgs boson mass obtained in these two
procedures are denoted $m^{p}_{H}$ and $m^{c}_{H}$, respectively.
An example of the two methods for determining $m_{H}$ is illustrated 
in Fig.~\ref{fig:fits}. We extract the fitted values $m_{H}^{p}$ and 
$m_{H}^{c}$ which agree within one standard deviation and both fits provide 
a suitable description of the data. 
The plots in this figure are for $m_{f} = 195$GeV. We note that we  
observe similar agreement between $m_{H}^{p}$ and $m_{H}^{c}$ for all
our choices of simulation parameters.  

\begin{figure}[H]
\begin{center}
\includegraphics[width=0.31\textwidth]{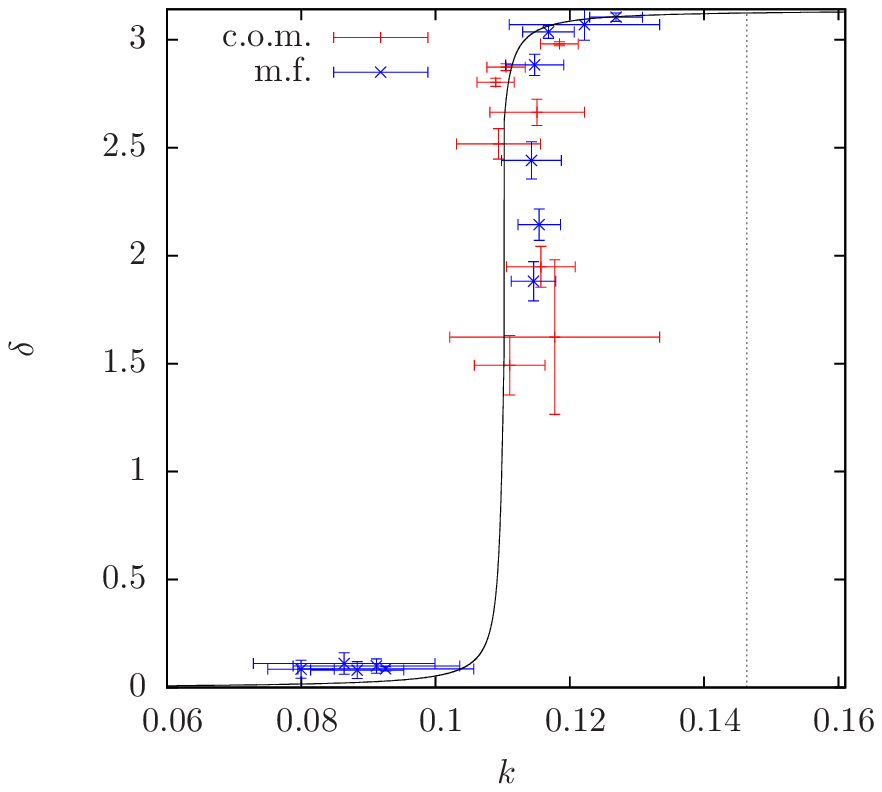}
\includegraphics[width=0.31\textwidth]{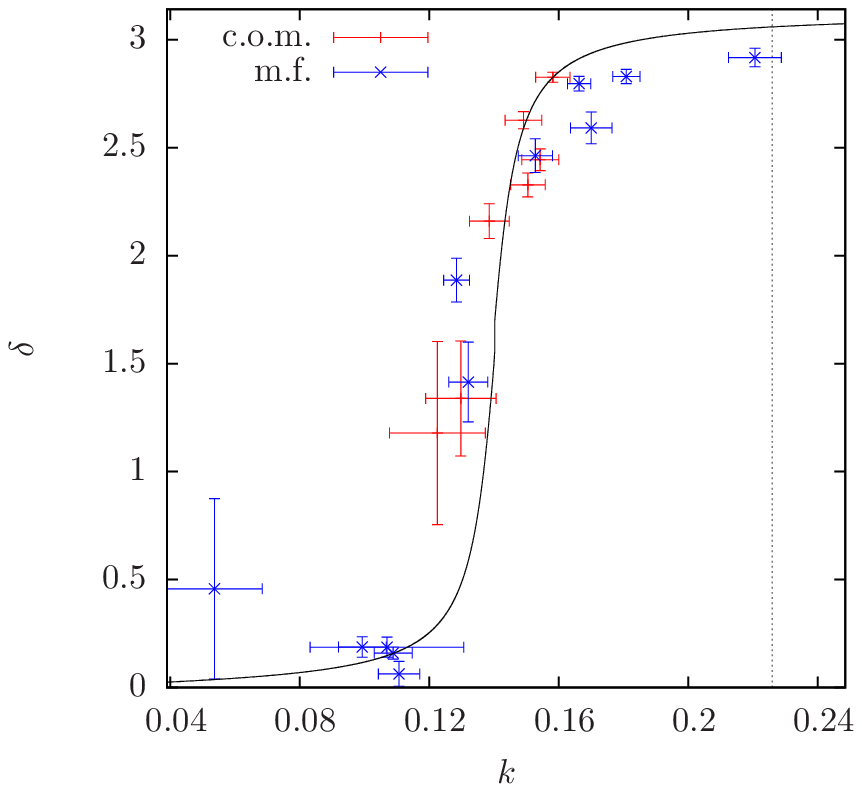}
\includegraphics[width=0.31\textwidth]{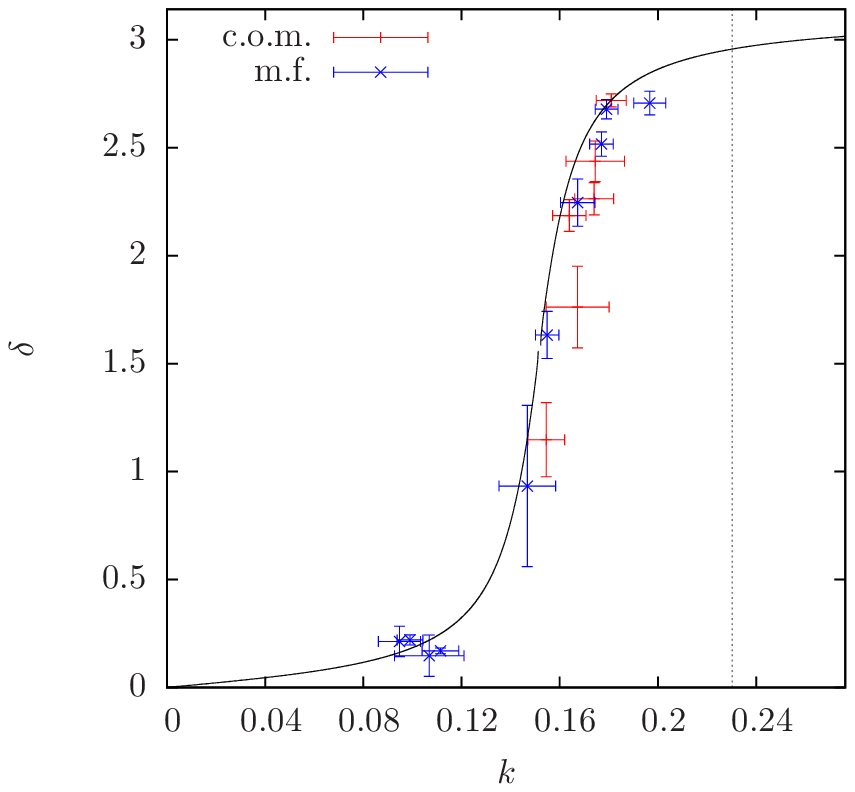}
\end{center}
\caption{Results for the scattering phase shifts at three values of $\hat{\lambda}$. 
From left to right, the plots correspond to $\hat{\lambda}=0.001, 10, \infty$, 
respectively. In each plot, the vertical dotted line indicates the 
position of the four-Goldstone threshold, above which our analysis method is 
inapplicable. Also, points obtained from both the centre of mass
system (c.o.m.)  and a system with one unit of total momentum (m.f.) are shown. 
Taken from Ref.~\cite{Gerhold:2011mx}.}
\label{fig:phases}
\end{figure}

To check the validity of the assumption that the Higgs boson is 
stable in our work,  a calculation of the Higgs 
boson resonance parameters has been performed in Ref.~\cite{Gerhold:2011mx}.
Since the finite volume techniques proposed in
Refs~\cite{Luscher:1990ux,Luscher:1991cf} are only applicable below
the inelastic threshold, external sources were introduced which give a mass to the
Goldstone bosons and break the O(4) symmetry explicitly.  
In the calculation the Goldstone 
boson energies were computed at non-zero momenta, using the original 
center of mass frame \cite{Luscher:1990ux,Luscher:1991cf} 
as well as a moving frame \cite{Rummukainen:1995vs,Feng:2010es}. 
By adjusting the values of the external source and the momenta, the 
Goldstone boson energies were tuned 
such that 
\beq
 2 E_{G} < m_{H} < 4 E_{G} \; .  
\eeq
The scattering phase shifts from
which the resonance 
parameters were extracted are shown in Fig.~\ref{fig:phases}, along with the 
position of the inelastic thresholds. These phase shifts are used to
fit the Breit-Wigner formula to determine the resonance mass and width.

The results of the Higgs boson width and mass obtained via the resonance 
analysis, perturbation theory, using the time-slice correlator and 
employing the momentum space Higgs boson propagator 
are shown in Tab.~\ref{table:comp}.  Here, the top quark 
mass has been set to its physical value. It is clear that 
the Higgs boson mass determined by the resonance study is 
consistent with that extracted from fits to the momentum space
propagator and the temporal correlation function. Furthermore, we see that at 
$m_f=m_t$, the width of the Higgs resonance is narrow, i.e. at 
most $\sim 10\%$ of the 
resonance mass in all cases.  From the results presented in this
table, it is demonstrated that it is justifiable to assume that the
Higgs boson width is zero since it turns out to be very narrow in the resonance 
analysis such that the width has no effect on the mass extraction.
\begin{table}[H]
\begin{center}
\begin{tabular}{|c|c|c|c|c|c|c|c|}\hline
  $\hat \lambda$&
   $\Lambda$ [GEV]&
    $m_H^{\rm resonance}$ &
      $\Gamma_H^{\rm resonance}$ &
         $\Gamma_H^{\text{pert}}$ &
            $m_H^p$ &
               $m_H^c$ \\\hline\hline
  $0.01$   & $883(1)$ & $0.278(3)$ & $0.0018(14)$ & $0.0054(1)$ & $0.278(2)$   & $0.274(4)$ \\\hline
  $1.0$    & $1503(5)$ & $0.383(6) $ & $0.0169(4)$ & $0.036(8)$ & $0.386(28)$ & $0.372(4)$ \\\hline
  $\infty$ & $1598(2)$ & $0.403(6) $ & $0.037(9)$ & $0.052(2)$ & $0.405(4)$   & $0.403(7)$ \\\hline
\end{tabular}
\end{center}
\caption{The results (taken from Ref.~\cite{Gerhold:2011mx}) of a 
study comparing the resonance parameters of the 
Higgs boson with the results of fits to the temporal correlation function and 
momentum space Higgs boson propagator. 
Errors are statistical only. Except for the cut-off scale, all
the results are in lattice units. The fermion mass is set to be the
physical top-quark mass. Results from three values of the quartic
coupling are presented. Also shown are the resonance mass and width 
from Breit-Wigner fits to the scattering cross-section. Finally, a 
perturbative estimate of the resonance width is
included. We note that because of some data losses the error on $m^{p}_{H}$ at
$\hat{\lambda}=1.0$ is larger than for the other parameters.}
\label{table:comp}
\end{table}

\subsection{Results of the Higgs boson mass bounds}
\label{sec:mass_bound_results}
\begin{figure}[H]
\begin{center}
\includegraphics[width=0.4\textwidth]{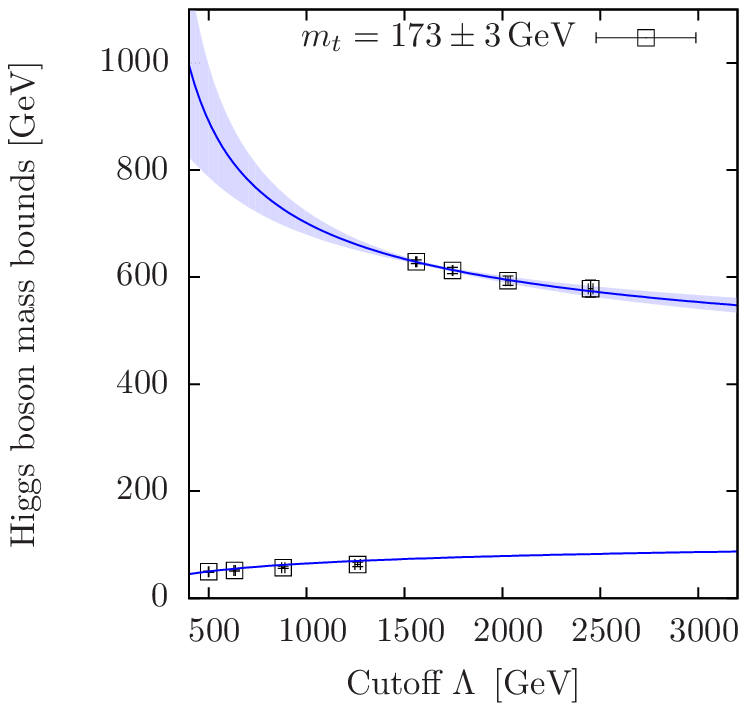}
\includegraphics[width=0.4\textwidth]{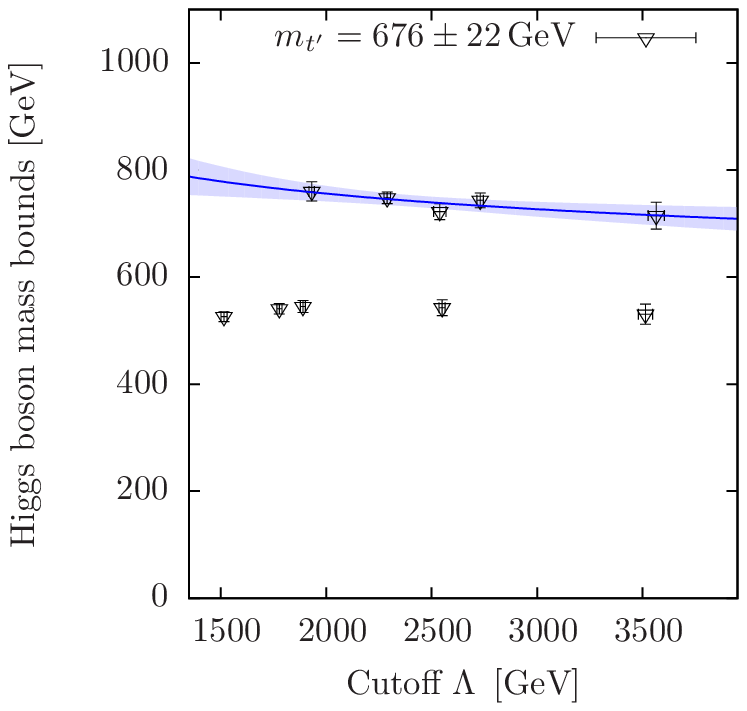} 
\end{center}
 \caption{The cut-off dependence of the upper and lower Higgs boson 
	mass bounds for fermion mass at $\sim 173$GeV (left) and
        $\sim676$GeV (right).  All data have been 
	extrapolated to infinite volume.} 
\label{fig:mass_bounds}
\end{figure}

We now turn to the results of the Higgs boson mass bound calculations 
discussed in the previous section. We first discuss the results of 
Ref.~\cite{Gerhold:2010bh}, where the upper and lower bounds were 
computed at several choices of the cut-off scale, with the fermion
masses at the physical top-quark mass, and also at $m_f \sim 676$GeV.
The main result from 
Ref.~\cite{Gerhold:2010bh} is shown in Fig.~\ref{fig:mass_bounds}. In the left 
graph, the situation for a SM top quark mass is shown. 
The right graph shows the situation for a fermion mass of $m_f \sim 676$GeV.
It can be clearly 
seen that while the upper bound is relatively unaffected when using a heavy 
fermion mass,
the lower bound increases substantially. 

\begin{figure}[H]
\begin{center}
\includegraphics[width=0.48\textwidth]{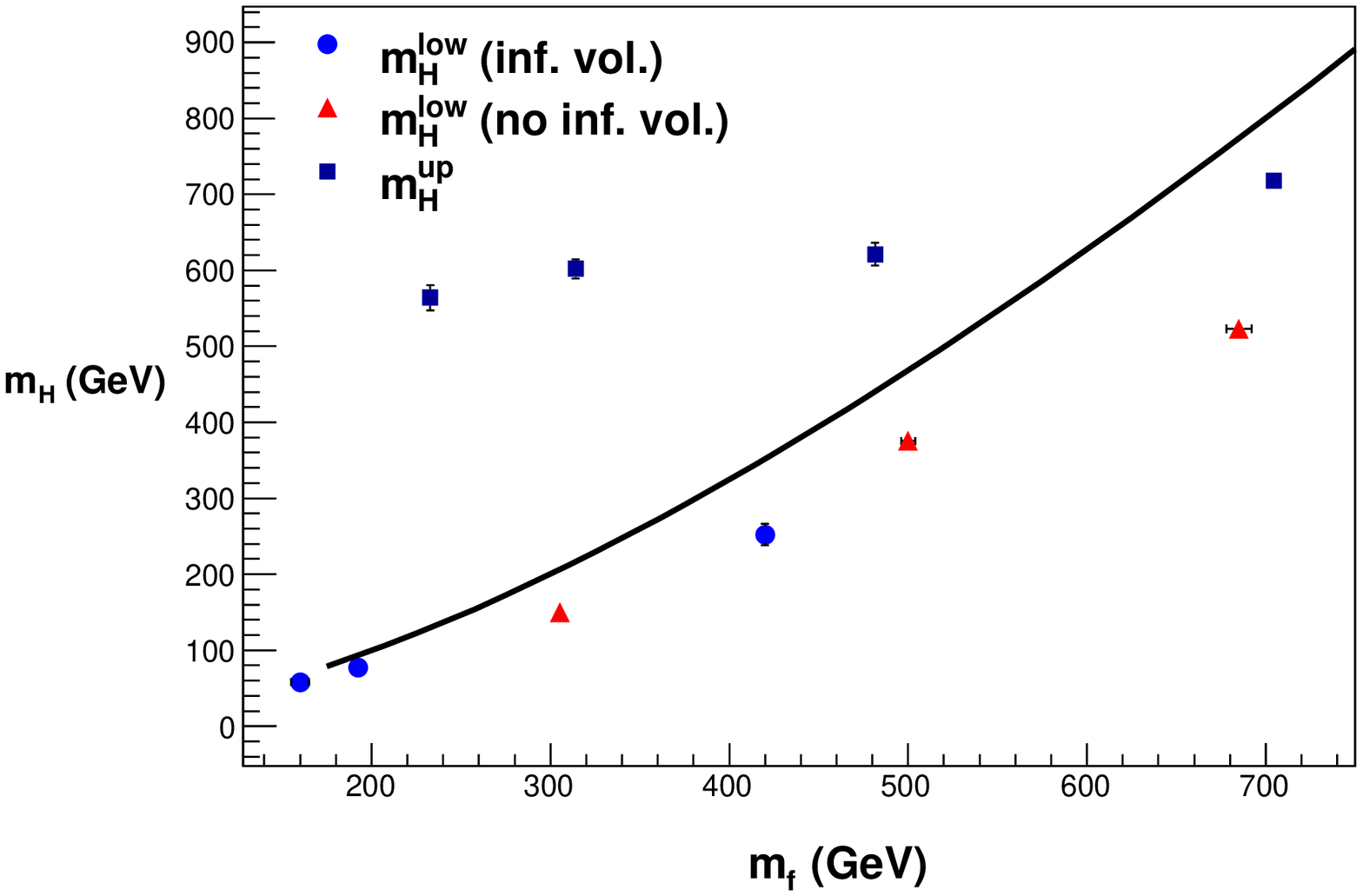}
\includegraphics[width=0.48\textwidth]{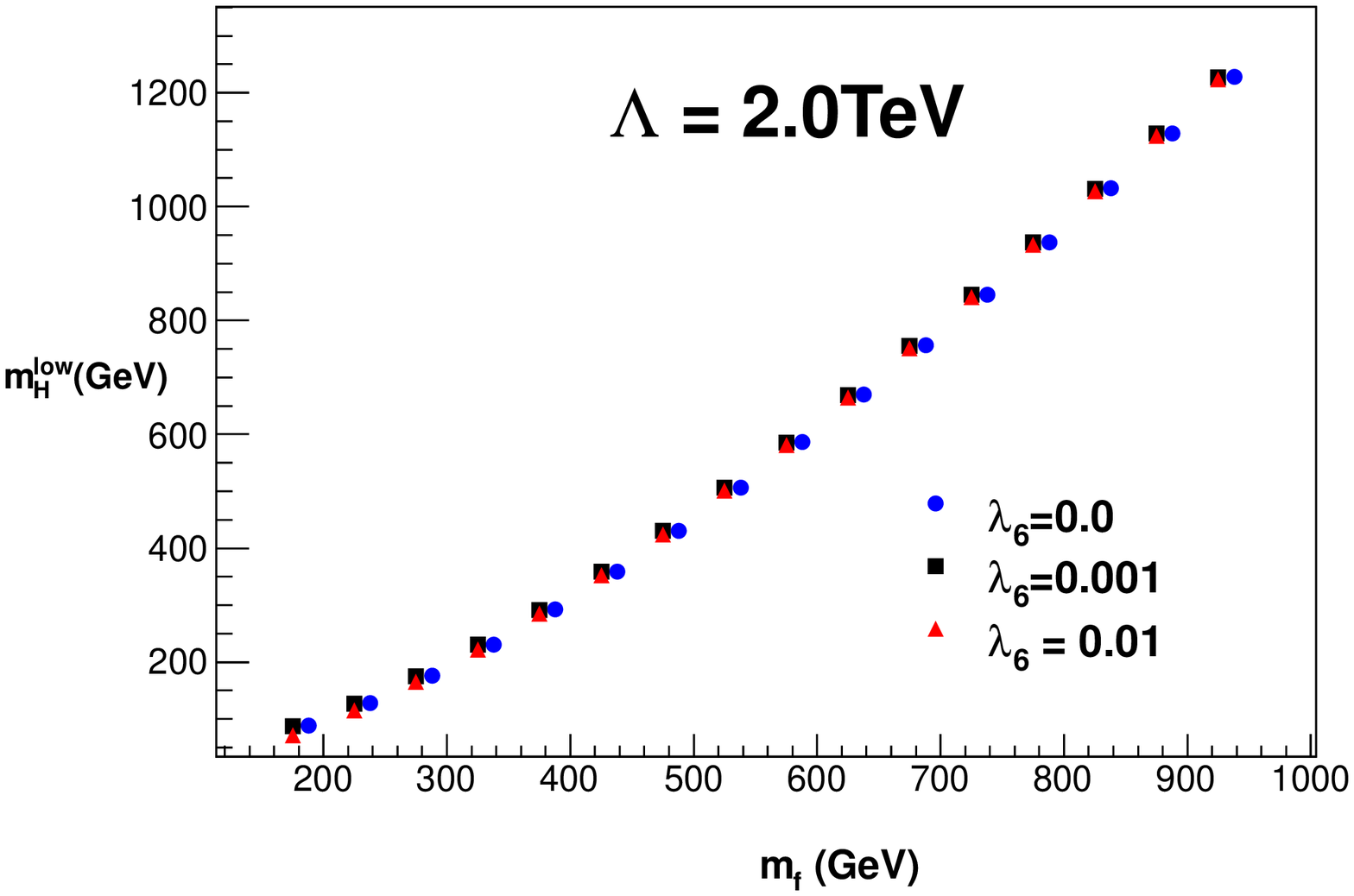}
\end{center}
\caption{Left: The dependence on the fermion mass of the upper and lower
  Higgs boson mass bounds, at the cut-off scale $\Lambda = 1.5$TeV.   
  Data points from lattice calculations are
  shown.  Results for the lower bound without infinite-volume
  extrapolation, using only $24^{3} \times 48$ lattices, are also 
  shown for comparison. The solid line results from 
  a one-loop calculation of the effective potential, as explained
  in the text.  Right: effects of a $\phi^{6}$ operator with 
  coupling $\lambda_{6}$ for the lower
  bound of the Higgs boson mass, at various fermion masses and the
  cut-off scale $\Lambda = 2$TeV.  Three values of the coupling
  constant $\lambda_{6}$ are plotted.}
\label{fig:mb_pert}
\end{figure}

Apart from the cut-off dependence of the bounds at a fixed value of $m_f$, the 
dependence of the bounds on $m_f$ itself has also been examined at a fixed 
value of the lattice cut-off~\cite{Bulava:2011ss}, the results of which are 
shown in Fig.~\ref{fig:mb_pert} (left). We clearly observe the 
increase of the lower bound with 
increasing $m_f$ in this figure. 
In particular,
Fig.~\ref{fig:mb_pert} suggests that with a Higgs boson mass of
$\sim125\mathrm{GeV}$, 
the mass of a mass-degenerate fourth generation of quarks is
restricted to be less than $\sim 350\mathrm{GeV}$. This is clearly already
below the bounds from direct experimental searches.

In addition to the numerical results, Fig.~\ref{fig:mb_pert} also contains the 
estimate of the lower bound from an effective 
potential calculation, which was performed 
using the same lattice regularisation as in our Monte Carlo 
simulation. In this calculation, the effective potential was
computed to one-loop order in the large${-}N_{f}$ limit.
Operationally, the one-loop calculations were carried out
by numerically computing the required momentum-mode summations in a
series of finite lattice volumes, and then extrapolating to the 
infinite-volume limit.  From this one-loop effective potential, $V$, 
the Higgs boson mass is determined by
solving for the scalar $vev$, $v$, and the Higgs boson mass in the 
gap equations,
\begin{align}
\label{eq:vacuum_stab}
\frac{d}{d \phi} V(\phi)|_{\phi=v} = 0, \quad
\frac{d^2}{d \phi^{2}} V(\phi)|_{\phi=v} = m_{H}^{2}\; .
\end{align} 
To compare to the numerically computed lower Higgs boson mass bound, 
in the effective potential calculation the 
quartic coupling has been set to zero. In addition, 
the cut-off and the fermion mass were fixed to the same 
values as in the simulations such that a direct comparison 
is possible. 
For a standard model top quark mass it has been demonstrated 
in \cite{Gerhold:2009ub,Gerhold:2010wy} that the lattice effective 
potential provides an excellent description for the numerical data
for the lower Higgs boson mass bound.

The left panel of Fig.~\ref{fig:mb_pert} clearly demonstrates that 
the trend of an increasingly higher value 
of the lower bound with increasing fermion masses, 
as suggested by the perturbative calculation is 
realised by the data up to very large values of $m_f$, although the 
quantitative agreement is better at low $m_f$.
Based on this qualitative agreement, we can examine the effect of
higher-dimensional operators in the effective potential using the same
loop and $1/N_{f}$ expansion.
To this end we include the contribution from the operator $\lambda_6\phi^6$ in 
the effective potential with 
$\lambda_{6}$ the coupling constant.
The addition of such an operator in the Lagrangian modifies the solution to
Eq.~(\ref{eq:vacuum_stab}), and can therefore alter the lower bounds on the
Higgs boson mass in principle.

Here we stress that the cut-off cannot be removed in the Higgs-Yukawa
model.  Furthermore, any perturbative expansion in this model is
only valid in the regime where the cut-off scale, 
$\Lambda = 1/a$,  is large enough compared to low-energy scales
such as the Higgs boson mass and the scalar $vev$.  In
Ref.~\cite{Luscher:1988uq}, it was demonstrated that $m/\Lambda < 0.5$ 
(with $m$ being a typical low-energy scale) is enough to ensure the
applicability of perturbation theory to the pure $\phi^{4}$ scalar
field theory. Here we impose the same condition, but 
on the value of the scalar field, in our perturbative
calculation for the effective potential for the Higgs-Yukawa model
including the $\lambda_{6} \phi^{6}$ operator.  This results in the 
stability criterion 
\begin{align}
\frac{d^2}{d \phi^{2}}V(\phi) > 0 , \quad \phi < 0.5 ,
\end{align}
where $\phi$ has been properly rescaled to be in lattice units.

In the right panel of Fig.~\ref{fig:mb_pert}, we show the results of
our investigation of the lower bounds on the Higgs boson mass, using
the one-loop effective potential including the contribution from the 
$\lambda_{6} \phi^{6}$ operator.
It is clear that in the regime where the perturbative expansion is valid,
a wide range of values of $\lambda_{6}$ lead to qualitatively very 
similar results.  Finally, we also point out that exploratory numerical Monte 
Carlo simulations which include the $\phi^6$ operator
agree with the perturbative results for a large range of bare 
Yukawa couplings~\cite{Gerhold:2010wy}.

\section{Study of the phase structure}\label{sec:phase_structure_results}
\subsection{Purposes and strategy of the study}
\label{sec:phase_transition_strategy}
It is an important task to explore the phase structure of the
Higgs-Yukawa model to identify the phase structure of the theory 
and determine the critical coupling constant values where a 
continuum limit can be performed. 
In this section, we will discuss two aspects concerning the phase structure
of the Higgs-Yukawa model considered here. 
The first are 
the locations of second-order bulk phase transitions in
the bare parameter space which can be identified as the continuum limits of
the lattice theory.   
For weak values of the bare Yukawa coupling the phase structure
has been investigated in \cite{Gerhold:2007yb,Gerhold:2007gx} and its 
knowledge was very helpful to identify the simulation parameters 
for the desired physical situation, i.e. a fixed value of the 
cut-off and the physical values of the fermion masses.
Here we remark that
the bounds on the Higgs boson and fermion masses as presented in 
Fig.~\ref{fig:mb_pert} in Sec.~\ref{sec:mass_bound_results} are
obtained in this weak bare Yukawa
coupling regime.  
In this section, we focus now on the 
large bare Yukawa coupling region and explore the phase structure
of the theory in this regime of the parameter space. 
The aim is to investigate, whether the phase transitions at large 
bare Yukawa coupling are governed by the same, Gaussian fixed point
as at small Yukawa coupling. If we would find deviations from 
the Gaussian fixed point behaviour, this would open the 
possibility  
that the renormalised Yukawa coupling can remain
strong up to a high cut-off scale which could lead 
to heavy fermion masses and even the existence of 
bound states.
We
have therefore been performing simulations at large values of bare Yukawa
coupling\footnote{In Ref.~\cite{Abada:1990ds,Hasenfratz:1991it}, 
it was demonstrated that in the
limit where the bare Yukawa coupling becomes infinity, the
Higgs-Yukawa model is equivalent to the pure O(4) scalar model.
However, our simulations are performed away from this limit.}, and
the exploratory results will be presented in Sec.~\ref{sec:bulk_phase_transition}.   
As a second aspect, we will present an 
investigation of the finite-temperature phase transition in
understanding the role of, in particular, heavy fermion masses
for the
electroweak phase transition, especially with respect to questions 
concerning baryogenesis~\cite{Cohen:1993nk}. 

Before detailing our on-going studies of the bulk and thermal phase transitions
of the Higgs-Yukawa model in the following two sections, here we describe
the general strategy in this work.

It is natural to use the $vev$ of the scalar
field to probe the phase structure.   However, a naive computation of
this $vev$ will always lead to vanishing results in lattice
calculations even in the broken phase, because of the 
finite volume as used in the simulations.  As discussed in the beginning of 
Sec.~\ref{sec:basic_observables},  it is appropriate to replace the 
scalar $vev$ with the magnetisation as defined in
Eqs.~(\ref{eq:bare_vev_definition}) and (\ref{eq:mag}).

In order to probe the nature of phase transitions, we have to determine
anomalous dimensions of the operators allowed by the symmetries. 
In finite
volume, second-order phase transitions are washed out and become
cross-overs, and the correlation length cannot exceed the size of the
system.   Therefore, for the study of the phase structure, we resort to finite-size
scaling techniques.  These techniques were developed originally by solving the 
renormalisation group equation (RGE) for finite-volume lattice systems
in condensed matter
physics~\cite{Fisher:1972zza}.  To draw analogy between field theory
and statistical mechanics, we also refer to these anomalous dimensions by
calling them critical exponents in this article, as usually done in statistical 
mechanics.

It is challenging to determine the anomalous dimension of the
operator corresponding to the Yukawa coupling term, because of the presence of
fermions and the flavour-changing structure of the operator.  
We will postpone the discussion of this operator
for future reports.  Here we focus on critical exponents in the scalar
sector.   To start, we calculate the susceptibility,
\begin{equation}
\label{eq:susc_def}
	\chi_{m} = V_{4} \left( \left< m^2\right> - \left< m\right>^2
        \right) ,
\end{equation}
which is the connected two-point function in the scalar sector.  This
quantity is proportional to the square of the correlation length,
$\xi$, and diverges at second-order phase transitions in the
infinite-volume limit.
Solving the RGE for this correlator for a finite-size system at fixed
cut-off scale (lattice spacing) near a second-order phase transition, 
one obtains the scaling law,
\begin{equation}
\label{eq:susc_scaling}
	\chi_{m}\left(t, L_s\right)\cdot L_s^{-\gamma/\nu} =  g\left(t
          L_s^{1/\nu}\right)\text{,\, with \,} t=\left (
          T/T_{c}  -1 \right )
\end{equation}
where $g$ is a universal scaling function, $L_s$ is the spatial
extent of the lattice, and $T_{c}$ is the critical temperature in the infinite-volume
limit, which could also be represented by the critical value of a particular
coupling.  The critical exponents, $\gamma$ and $\nu$ are related to the
anomalous dimensions of the scalar field and the mass operator,
$\phi^{2}$.   This scaling behaviour is exact near the critical point
for space-time dimension, $d < 3$.   Therefore it is an appropriate
tool in our study of the finite-temperature phase transition.
However, in the investigation of the bulk phase structure, 
we have a $d=4$ field theory, and the above scaling
relation should be modified because of
triviality~\cite{Brezin:1981gm,Brezin:1985xx,Bernreuther:1987hv,Kenna:1992np,Kenna:2004cm}, 
if the transition is governed by a Gaussian fixed point.  These
modifications appear as logarithmic corrections in
Eq.~(\ref{eq:susc_scaling}).  They are not included in the analysis
presented in this article, but are being considered in our on-going work.
As will be discussed in the following, the scaling tests and
the extraction of anomalous dimensions using
Eq.~(\ref{eq:susc_scaling}) are complicated because of the number of
free parameters that are involved in the methods for modelling the
unknown universal function, $g$.  In particular, it is difficult to accurately
determine $\nu$ using this procedure.  This complication can be reduced by
studying Binder's cumulant~\cite{Binder:1981sa},
\begin{equation}
\label{eq:binders_cumulant_def}
	Q_{L} = 1 - \frac{\left< m^4\right>}{3\left< m^2\right>^{2}} .
\end{equation}
This quantity is simply the connected four-point function, normalised
by the square of the two-point function, in the scalar sector.  Because
of the normalisation, $Q_{L}$ is independent of the critical exponent
$\gamma$.  Furthermore, it is related to the renormalised scalar
quartic coupling in the infinite-volume limit by a proportionality factor 
$V_{4}/\xi^{4}$~\cite{Freedman:1982zu}.    Since Binder's cumulant is
normalised to be dimensionless,  its values computed on different 
(dimensionless) lattice sizes with the same cut-off scale will coincide with each other at the
critical point.  It is also expected to exhibit milder
scaling violations resulting from higher-dimensional
operators~\cite{Privman:1983dj, Binder:1985FSTH}.

In the next three sections, we discuss details of the investigation of the thermal
and bulk phase structures using the quantities defined in this
section.  Errors on all the numerical results in this section are
statistical only.

\subsection{Bulk phase structure at small Yukawa
  couplings}\label{sec:bulk_phase_weak_y}
Before reporting the details of our on-going investigation in the bulk
phase structure of the Higgs-Yukawa model in the strong-Yukawa regime,
we briefly summarise the results obtained in the region of weak-Yukawa 
coupling~\cite{Gerhold:2007gx} in this section. The order
parameters characterising the different phases are the magnetisation
defined in Eqs.~(\ref{eq:bare_vev_definition}) and (\ref{eq:mag}), 
and the staggered magnetisation
\begin{equation}
 \langle s \rangle  = \left\langle  \frac{1}{V_{4}}
   \sum_{x} (-1)^{x_{1}+x_{2}+x_{3}+x_{4}}  \left ( \sum_{\alpha}
 |\phi_{x}^{\alpha}|^{2} \right )^{1/2}  \right\rangle \; .
\end{equation}
The staggered magnetisation is relevant for the breaking of the
symmetry,
\bea
  \kappa &\longrightarrow&  -\kappa ,\nonumber\\
\label{eq:staggered_symmetry}
  \phi^{\alpha}_{x} &\longrightarrow& (-1)^{x_{1}+x_{2}+x_{3}+x_{4}}
  \phi_{x}^{\alpha} ,
\eea
in the action in Eq.~(\ref{eq:action_lattice_bosonic}).  

In the Higgs-Yukawa model, four phases have been observed:
\begin{enumerate}
 \item  A symmetric (SYM) phase with $\langle m \rangle = \langle s
   \rangle = 0$.
 \item  A broken, or ferromagnetic (FM), phase with $\langle m \rangle
   \not= 0$ but $\langle s \rangle = 0$.
 \item  A staggered-broken, or anti-ferromagnetic (AFM), phase with $\langle m \rangle
   = 0$ but $\langle s \rangle \not= 0$.
 \item  A ferrimagnetic (PI) phase with $\langle m \rangle
   \not= 0$ and $\langle s \rangle \not= 0$.
\end{enumerate}

\begin{figure}[H]
	\begin{center}
	$	
	\begin{array}{ccc}
		\hspace*{-0.1cm} \includegraphics[width=0.33\linewidth]{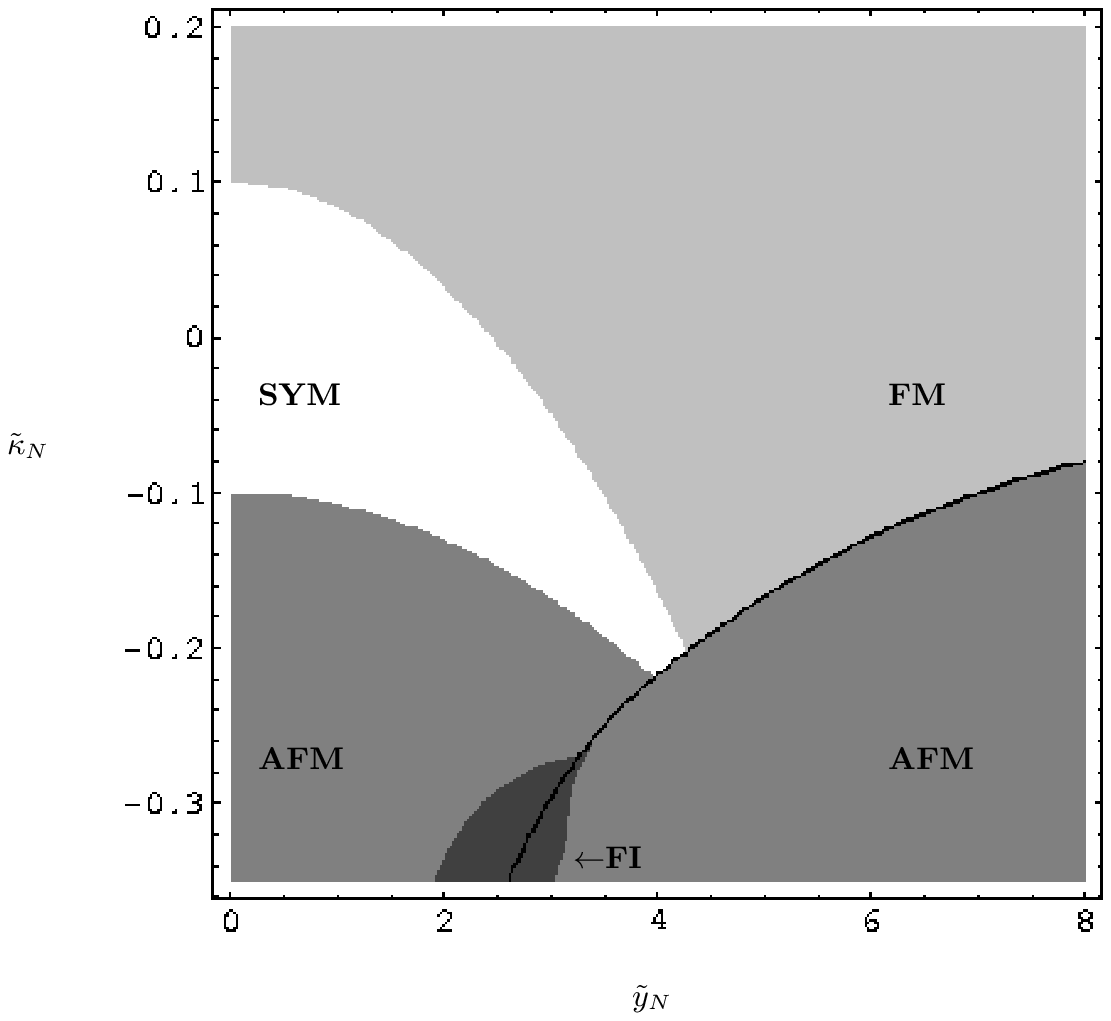}  &
		\hspace*{-0.15cm} \includegraphics[width=0.33\linewidth]{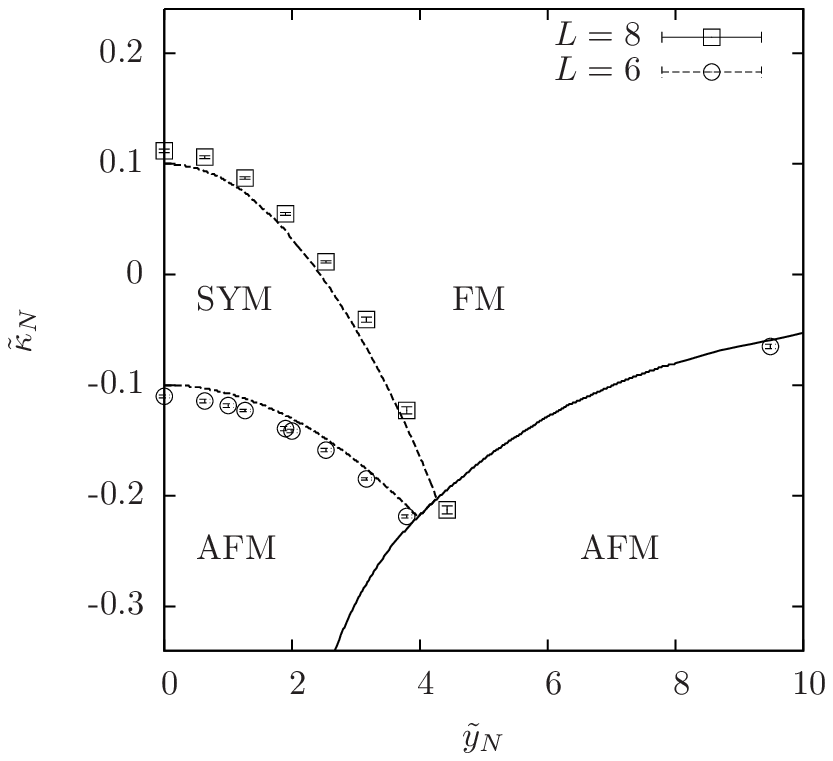} &
		\hspace*{-0.3cm} \includegraphics[width=0.33\linewidth]{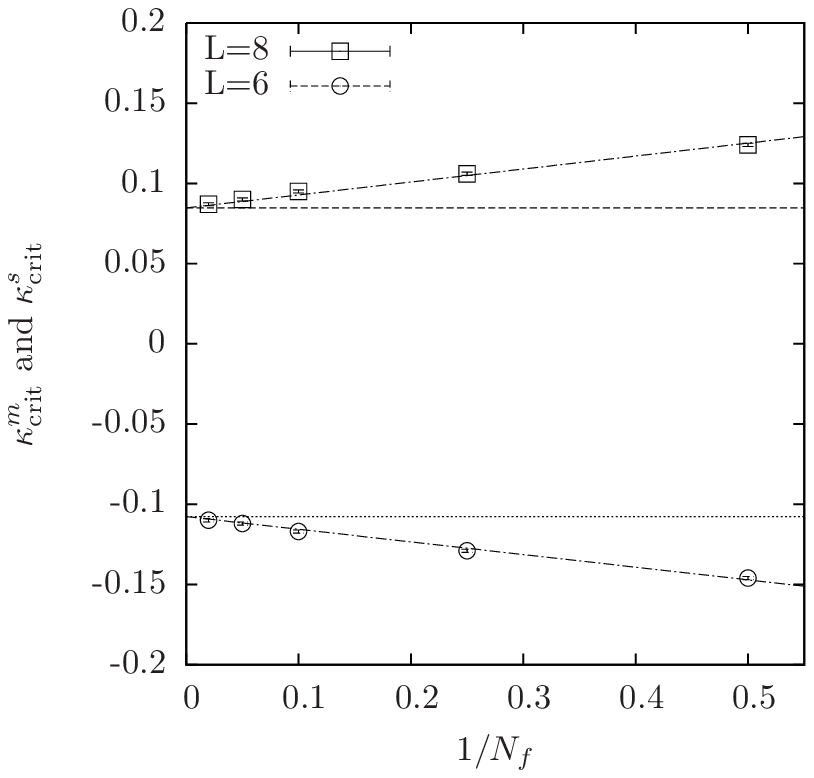}
	\end{array}
	$	
	\end{center}
\vspace*{-0.5cm}
\caption{The phase structure analysis. The left panel shows analytical predictions for the case of $L_s=\infty$, $N_f = 
\infty$ and $\tilde \lambda_N = 0.1$. The black line indicates a first order phase transition, while all other transitions are of 
second order. The middle panel demonstrates a numerical test of the
transitions from the SYM to both FM and AFM phases with 
$N_f=10$. The right panel displays the $N_f$ dependence in the
critical values of $\kappa$ for the SYM${-}$FM and SYM${-}$AFM
transitions, at $\tilde \lambda_N = 0.1$ and $\tilde y_N = 1.0$.
These critical $\kappa$ values are denoted as $\kappa_{\rm crit}^{m}$
($> 0$)
and $\kappa_{\rm crit}^{s}$ ($< 0$),
respectively. The squares and circles in
the middle and right panels of the figures come from direct 
numerical simulations on the indicated lattice sizes.}
\label{fig:phaseDiagramSummary}
\end{figure}

Our current knowledge of the phase structure of the Higgs-Yukawa model
in the weak-Yukawa regime is summarised in Fig.~\ref{fig:phaseDiagramSummary}.
To make it convenient in comparing results from numerical simulations
to a large${-}N_{f}$ analytic calculation~\cite{Gerhold:2007yb}, we have performed the change
of variables,
\begin{equation}
 \hat y = \frac{\tilde y_N}{\sqrt{N_f}}, \quad \kappa =
 \tilde{\kappa}_{N}, \quad \hat \lambda = \frac{\tilde
   \lambda_N}{N_f}, \quad \Phi = \sqrt{N_f} \tilde \Phi ,
\end{equation}
in the plots in this figure.   The large${-}N_{f}$ calculation was
carried out in the $N_f \rightarrow \infty$ limit while keeping
$\tilde y_N$, $\tilde \lambda_N$ and $\tilde \Phi$ fixed.  The left
panel of Fig. ~\ref{fig:phaseDiagramSummary} is the result from the
large${-}N_{f}$ calculation, and the middle panel is the comparison
between this calculation and the numerical results from lattice
simulations at $N_{f} = 10$.  The right panel of this figure shows the $N_{f}$
dependence on the critical values of $\kappa$ at the SYM${-}$FM and 
FM${-}$AFM transitions in our numerical calculation, with the Yukawa
coupling set to $\tilde{y}_{N} = 0.1$.   It is observed that the
$N_{f}$ dependence appears to be mild.   This indicates that the
large${-}N_{f}$ analytic calculation may serve as a reasonable,
qualitative, guide in choosing the simulation parameters for the
numerical simulations. Although this analysis has been 
performed in the weak Yukawa coupling region, the good qualitative 
description makes it possible to also use the large
$N_f$ expansion also in the strong-Yukawa regime, which 
was indeed observed in \cite{Gerhold:2007yb}.
%

In the weak Yukawa coupling region, we concentrated on the study of
the SYM${-}$FM phase transition, which was confirmed to be
second-order.  This allowed us to study physically interesting
quantities, such as the Higgs boson mass bound presented in
Sec.~\ref{sec:higgs_mass_bounds}, near this phase-transition with good 
control of the cut-off dependence.
%

\subsection{Bulk phase transition at large Yukawa couplings}\label{sec:bulk_phase_transition}
It is not well understood how the Higgs-Yukawa model at large bare Yukawa couplings 
differs from that in the weak-coupling regime. A first step in a detailed 
analysis and hence a deeper understanding of the model in this region is the 
investigation of the bulk phase transitions. It can be shown that the 
Higgs-Yukawa model reduces to a pure scalar non-linear $\sigma$-model at 
infinite bare Yukawa couplings~\cite{Abada:1990ds,Hasenfratz:1991it}, and hence becomes 
trivial at a certain cut-off scale. However, it is not clear what 
happens at large but finite Yukawa couplings. To be able to detect any 
differences from a Gaussian (trivial) theory the critical exponents 
of the phase transition have to be extracted and compared with those 
of the O(4) model. If the strong-coupling regime is indeed different 
from the weak-coupling one and hence would be governed 
by a non-trivial fixed point\footnote{There has been early lattice work on the 3-dimensional 
Higgs-Yukawa model~\cite{Focht:1995ie}, attempting at finding fix points 
that are different from that of the pure scalar field theory.}, 
it would be very interesting to investigate 
the possibility of very heavy fermions which give rise to a fourth generation, 
while still maintaining a light Higgs boson in the theory. In such a scenario 
it is unclear, whether an analysis as, e.g. \cite{Eberhardt:2012gv} is applicable and 
also, whether the Higgs boson mass bounds of section III are valid. 

\begin{figure}[H]
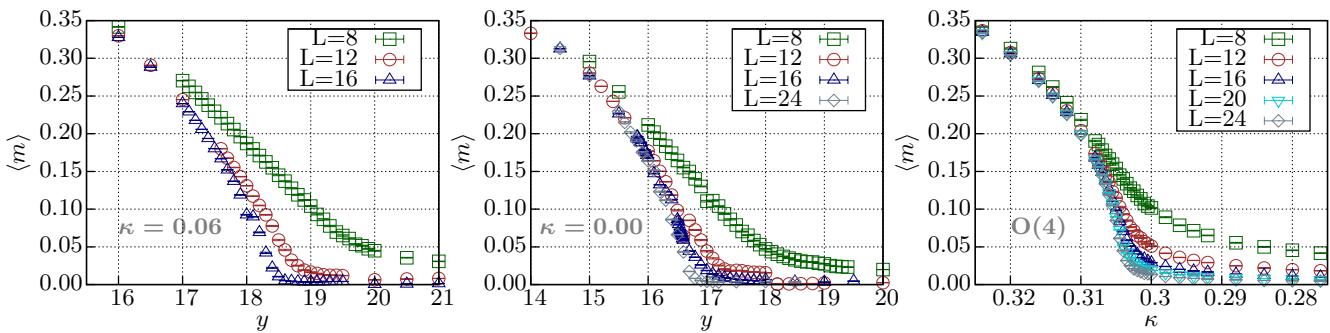

	\begin{center}
	$	
	\begin{array}{ccc}
		\hspace*{-0.5cm}  \input{./figures/StrongYukawaCoupling/mag_kap006.tex} &
		\hspace*{-1.2cm}  \input{./figures/StrongYukawaCoupling/mag_kap000.tex} &
		\hspace*{-1.2cm}  \input{./figures/StrongYukawaCoupling/mag_O4.tex} 
	\end{array}
	$	
	\end{center}
\vspace*{-0.5cm}
\caption[Magnetisation at large Yukawa couplings.]{Magnetisation, 
$\left< m\right>$, for the Higgs-Yukawa model at $\kappa=0.06$ (left), 
$\kappa=0.00$ (middle) and the pure O(4) model (right) for various volumes. 
For the O(4) $\left< m\right>$ is plotted as a function of {\em decreasing} $\kappa$ 
to match optically with the Higgs-Yukawa model. 
The absence of discontinuities in $\left< m\right>$ is an evidence for a second order phase transition.}	
\label{MagAtLargeY}
\end{figure}

The magnetisation, defined in Eqs.~(\ref{eq:bare_vev_definition}) and 
(\ref{eq:mag}), can act as an order parameter to identify and determine 
the order of the phase transition. In Fig.~\ref{MagAtLargeY}, the 
magnetisation for the Higgs-Yukawa model obtained on different lattice 
volumes is shown as a function of $y$ for two $\kappa$ values. In addition, 
we show the magnetisation as a function of $\kappa$ for the O(4) model. 
The SYM and FM phases can be clearly distinguished and the phase transition 
is washed out because of finite volume effects as previously discussed.

The absence of any discontinuities in the magnetisation is strong evidence for a second-order phase transition in all three depicted cases. In general, second-order phase transitions are classified through their critical exponents and the question arises if these exponents are different in the strong-Yukawa and pure O(4) models.  To answer this question, a careful investigation of the susceptibility and  Binder's cumulant will be presented in the following. 

\begin{figure}[H]
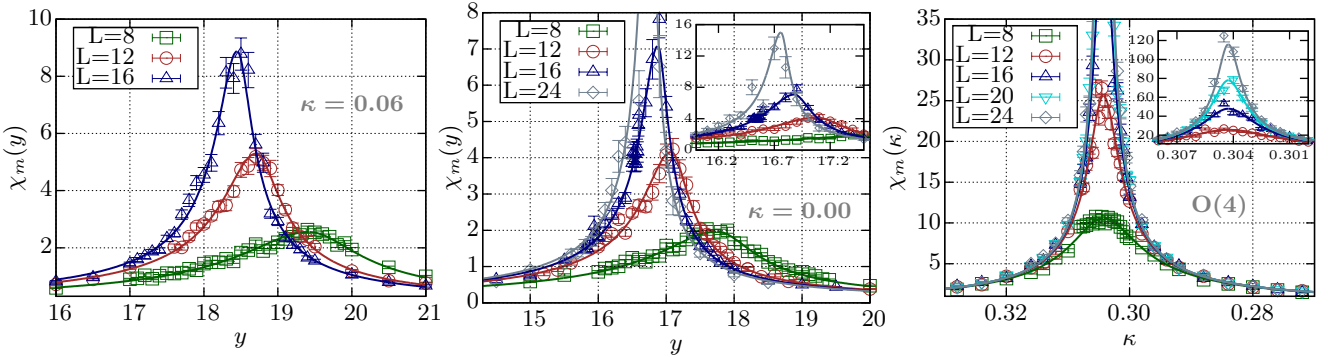

	\begin{center}
	$	
	\begin{array}{ccc}
		\hspace*{-0.5cm} \input{figures/StrongYukawaCoupling/susceptibility_kap006.tex} &
		\hspace*{-1.2cm} \input{figures/StrongYukawaCoupling/susceptibility_kap000.tex} &
		\hspace*{-1.2cm} \input{figures/StrongYukawaCoupling/susceptibility_O4.tex} 
	\end{array}
	$	
	\end{center}
\vspace*{-0.5cm}
\caption[Susceptibility at large Yukawa couplings.]{Susceptibility 
$\chi_m$ at $\kappa=0.06$ (left), $\kappa=0.00$ (middle) and the 
O(4) model (right) for various volumes. The curves are the result of a fit to 
Eq.~(\ref{FitFunction}). The right top boxes in the middle and the right 
panels show $\chi_m$ for the largest volumes. For the Higgs-Yukawa model a 
volume-dependent shift of $y_c$ towards $y_c^{(L=\infty)}$ can be observed.  
This shift is not observed in the O(4) model.}	
\label{SusAtLargeY}
\end{figure}

The critical exponents can be calculated by using the finite-size scaling of the 
susceptibility, Eq.~(\ref{eq:susc_def}). The susceptibility is 
shown in Fig.~\ref{SusAtLargeY} for the Higgs-Yukawa and O(4) models. 
This quantity diverges at the critical point in the infinite volume limit. 
Such a divergence in infinite volume is reflected in 
a bulk finite-size scaling behaviour in lattice calculations. 
As mentioned before in Eq.~(\ref{eq:susc_scaling}), the finite-size scaling is 
predicted by renormalisation group theory, with modifications resulting from 
scaling violation such as that discussed in Ref.~\cite{Fisher:1972zza},
\begin{equation}
	\chi_{m}\left(t, L\right)\cdot L_s^{-\gamma/\nu} =  g\left(\hat t L_s^{1/\nu}\right)\text{,\, with \,} \hat t=\left[T/\left(T_c^{(L=\infty)} - C\cdot L_s^{-b}\right) -1\right],
	\label{ScalingFunction}
\end{equation}
where $C$ is a phenomenological parameter and $b$ is a shift exponent~\cite{Fisher:1972zza}. 
This modification comes from the fact that the position of the maximum of 
$\chi_m$ is volume dependent. From Eq.~(\ref{eq:susc_scaling}) the infinite-volume 
critical temperature can be extracted directly. 
For the O(4) model we do not observe any shift of the maximum and 
hence Eq.~(\ref{eq:susc_scaling}) is a good description of our data in 
this case. It should be stressed, that the temperature, 
$T$, in this section is the control parameter.  
In our work, it is either the Yukawa coupling, $y$, in the Higgs-Yukawa model or the hopping parameter, $\kappa$, in the pure O(4) model.
To extract the critical exponents from the susceptibility, we perform a simultaneous fit of all data obtained at all volumes to the partly-empirical formula~\cite{Jansen:1989gd},
\begin{align}
	\chi_{m} = A\left(L_s^{-2/\nu} + B\left[T/T_c^{(L=\infty)} - C\cdot L_s^{-b} -1\right]^2\right)^{-\gamma/2}.
	\label{FitFunction}
\end{align}
This formula was also used for a fit to $\chi_m$ of the O(4) model, but with the 
modification of excluding the parameters $C$ and $b$ because of the reasons 
mentioned above. The fit results are summarised in Tab.~\ref{TableFitSus} 
and will be discussed later.  Notice that there may be logarithmic 
corrections to the scaling behaviour of the susceptibility because  
triviality may still be present also in the strong-Yukawa model.  
These corrections should, in principle, be included in 
Eq.~(\ref{FitFunction})\footnote{These logarithmic corrections are 
surely present in the finite-size scaling behaviour of the 
susceptibility in the pure O(4) 
model~\cite{Brezin:1981gm,Brezin:1985xx,Bernreuther:1987hv,Kenna:1992np,Kenna:2004cm}.  
However, our exploratory numerical results show that their inclusion produces 
minor changes in the results of the critical exponents in the O(4) model.}.  
This is on-going work, and the result will be presented in a later publication.
Therefore, we consider our present values of the critical exponents as preliminary 
and they should be taken with caution. 

\begin{table}[H]\centering{
	\begin{tabular}{|c||c|c|c|c|c||c|}\hline
		& $T_c^{(L=\infty)}$ & $\nu$ & $\gamma$ & $C$ & $b$ & fit interval \\ \hline\hline
		$\kappa=0.06$ & 18.119(67) & 0.576(28) & 1.038(30) & 4.7(1.6) & 1.95(18) & 17.5, 20.0 \\ \hline   
		$\kappa=0.00$ & 16.676(15) & 0.541(22) & 0.996(15) & 10(2) & 2.42(10) & 15.0, 19.0\\ \hline  
		O(4)                    & 0.304268(27) & 0.499(12) & 1.086(19) & N/A & N/A & 0.300, 0.308 \\ \hline
	\end{tabular}}
\caption[Fit results to Eq.~(\ref{FitFunction}).]{Results of a correlated fit to the susceptibility according to Eq.~(\ref{FitFunction}) where the last column indicates the fit interval. The parameter $T$ stands either for $y$ in the Higgs-Yukawa model or for $\kappa$ in the O(4) model. Since no volume-dependent shift can be observed in the O(4) model for $\chi_m$, the parameters $C$ and $b$ have not been fitted here. All quoted errors are statistical only.}	
\label{TableFitSus}
\end{table}

\begin{figure}[H]
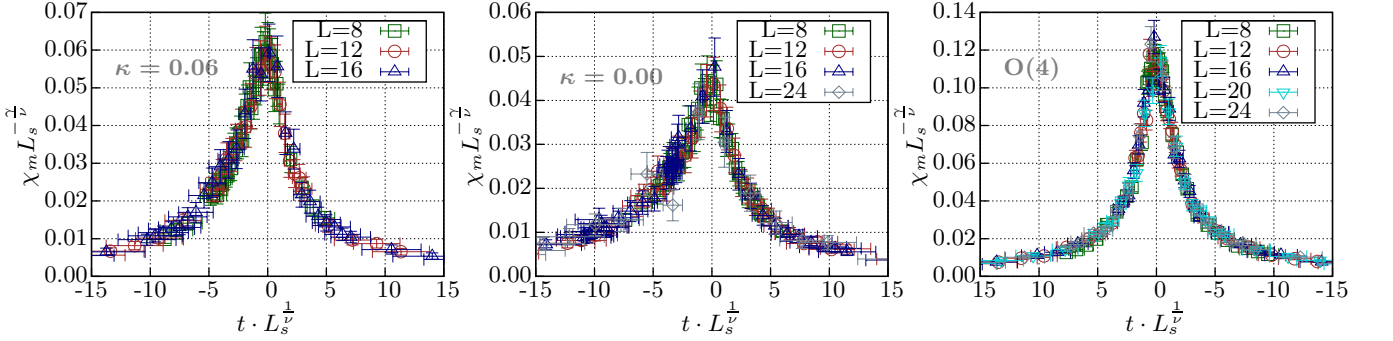

	\begin{center}
	$	
	\begin{array}{ccc}
		\hspace*{-0.5cm}  \input{figures/StrongYukawaCoupling/sus_rescaled_kap006.tex} &
		\hspace*{-1.2cm}  \input{figures/StrongYukawaCoupling/sus_rescaled_kap000.tex} &
		\hspace*{-1.2cm}  \input{figures/StrongYukawaCoupling/sus_rescaled_O4.tex} 
	\end{array}
	$	
	\end{center}
\vspace*{-0.5cm}
\caption[Rescaled susceptibility at large Yukawa couplings.]{Scaling behaviour of susceptibility at $\kappa=0.06$ (left), $\kappa=0.00$ (middle) and the O(4) model (right) for various volumes.}
\label{SusRescaledAtLargeY}
\end{figure}

It is possible to re-scale the susceptibility according to Eq.~(\ref{ScalingFunction}) 
for the Higgs-Yukawa theory, or Eq.~(\ref{eq:susc_scaling}) for the O(4) model, 
respectively. The fitted parameters extracted from Eq.~(\ref{FitFunction}) 
can be used to construct $\chi_{m}\left(t, L_s\right)\cdot L_s^{-\gamma/\nu}$ and 
test its scaling against $t\cdot L_s^{1/\nu}$. This is shown in 
Fig.~\ref{SusRescaledAtLargeY}. Points for all volumes collapse on the 
same curve in each of the three cases shown. This behaviour is typical for 
second-order phase transitions and hence provides further evidence 
that such a second-order transition happens in the regime of strong Yukawa couplings.

\begin{figure}[H]
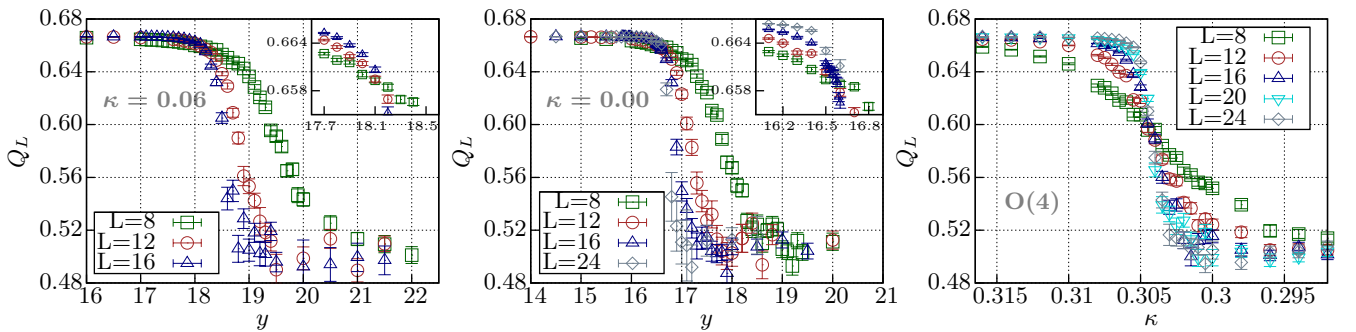

	\begin{center}
	$	
	\begin{array}{ccc}
		\hspace*{-0.5cm}  \input{figures/StrongYukawaCoupling/binder_kap006.tex} &
		\hspace*{-1.2cm}  \input{figures/StrongYukawaCoupling/binder_kap000.tex} &
		\hspace*{-1.2cm}  \input{figures/StrongYukawaCoupling/binder_O4.tex} 
	\end{array}
	$	
	\end{center}
\vspace*{-0.5cm}
\caption[Binder cumulant at large Yukawa couplings.]{Binder's Cumulant $Q_L$ at $\kappa=0.06$ (left), $\kappa=0.00$ (middle) and the O(4) model (right) for various volumes where the subscript L indicates the finite volume quantity. 
Note that the value of $Q_L$ at the critical point is different in the Higgs-Yukawa and the O(4) models.}	
\label{BinderAtLargeY}
\end{figure}

An alternative way of determining critical exponents is via Binder's cumulant, 
Eq.~(\ref{eq:binders_cumulant_def}). One advantage of this quantity over the susceptibility is its milder power-law scaling violation which is given by 
\begin{equation}
\label{eq:Binder_cumulant_scaling}
	Q_L  = g_{Q_L}\left( tL^{1/\nu}\right),
\end{equation}
where $g_{Q_L}$ is a universal function and $t$ is defined in Eq.~(\ref{eq:susc_scaling}). This behaviour can be observed in Fig. \ref{BinderAtLargeY} where all volumes intersect at the phase transition point in infinite volume where $t=0$. Even for the Higgs-Yukawa model no shift can be observed and hence the parameters $C$ and $b$ can be completely neglected
in the scaling variable.

The value of Binder's cumulant in the broken phase comes from the fact that 
$\left< m^4\right>\approx\left< m^2\right>^2$ and hence $Q_L\approx2/3$ \cite{Binder:1981sa}. 
Our results for $Q_L$ at the critical point come close to this value for all
setups considered here. Still, $Q_L$ obtained in the Higgs-Yukawa model differs 
from the one in the O(4) model. This may arise from effects of finite renormalisation because of the inclusion of fermions. Its implication in the difference of the O(4) model and the Higgs-Yukawa model is under investigation now.  Furthermore, it can be demonstrated that for Binder's cumulant, as contrary to the susceptibility,  there is no logarithmic corrections to the scaling behaviour arising from triviality in the pure O(4) model~\cite{Bernreuther:1987hv}.  Whether or not such corrections can be present in the Higgs-Yukawa model is being studied now.  

The basic idea of extracting the critical exponent, $\nu$, from Binder's cumulant is the use of the curve collapse of Eq.~(\ref{ScalingFunction}). If the scaling function $g_{Q_L}$ is known one will simply minimise \cite{Bhattacharjee:2001MDCS}
\begin{equation}
	R_{Q_L} =\frac{1}{N}\sum\left| Q_L - g_{Q_L}\left(tL^{1/\nu}\right)\right|, \mbox{ } (N = {\rm total}\mbox{ }{\rm number}\mbox{ }{\rm of}\mbox{ }{\rm data}\mbox{ }{\rm points})
	\label{BinderMinimizationWithKnowng}
\end{equation}
which would allow to extract $\nu$ as a direct consequence of the scaling behaviour. The sum is taken over all data points, and $R_{Q_L}$ is minimal for the correct choice of the parameters $\nu$ and $T_c^{L=\infty}$. In the absence of any statistical and systematic errors the function $R_{Q_L}$ would become zero. 

\begin{figure}[H]
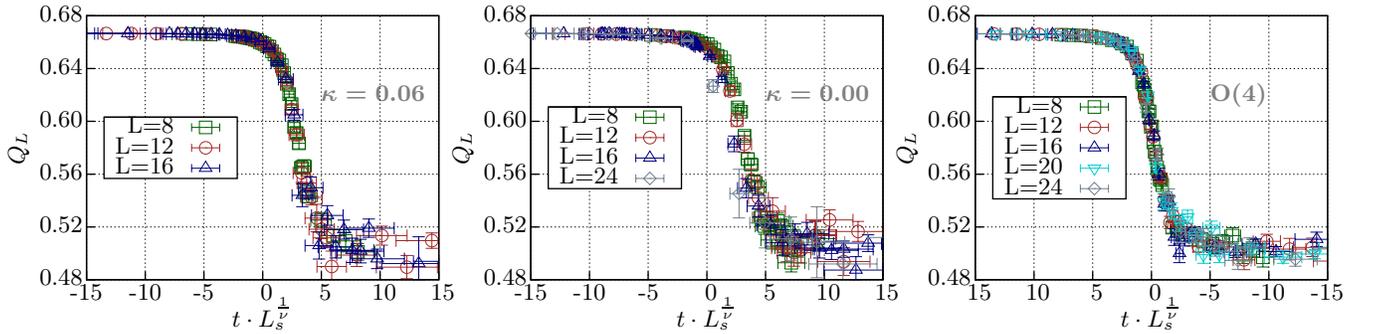

	\begin{center}
	$	
	\begin{array}{ccc}
		\hspace*{-0.5cm}  \input{figures/StrongYukawaCoupling/binder_rescaled_kap006.tex} &
		\hspace*{-1.2cm}  \input{figures/StrongYukawaCoupling/binder_rescaled_kap000.tex} &
		\hspace*{-1.2cm}  \input{figures/StrongYukawaCoupling/binder_rescaled_O4.tex} 
	\end{array}
	$	
	\end{center}
\vspace*{-0.5cm}
\caption[Scaling behaviour of Binder cumulant at large Yukawa couplings.]{Scaling behaviour of 
Binder's cumulant at $\kappa=0.06$ (left), $\kappa=0.00$ (middle) and the 
O(4) model (right) for various volumes using the parameters listed in table~\ref{TablePBmini}.}	
\label{BinderRescaledAtLargeY}
\end{figure}

The scaling function $g_{Q_L}$ is unknown. However, this can be overcome by the observation 
that any volume, in the following called $p$, can act as a reference function for the 
correct choice of parameters, taking thus over the role of 
$g_{Q_L}$. Instead of minimising Eq.~(\ref{BinderMinimizationWithKnowng}), we minimise \cite{Bhattacharjee:2001MDCS}
\begin{equation}
	P_b = \left[ \frac{1}{N_\text{over}}\sum_p\sum_{j\ne p}\sum_{i, \text{over}} \left| Q_{L_j}- \mathcal{E}_p\left( t_{ij} L_j^{1/\nu} \right) \right|^2 \right]^{1/2}.
\end{equation}
Here, the scaling function is replaced by the interpolating function $\mathcal{E}_p$ which is constructed by interpolating the data points obtained on volume $p$ to volume $j$ for the values of the scaling variable $t_{ij} L_j^{1/\nu}$ , with the index $i$ going through all data points of volume $j$. In our case, $\mathcal{E}_p$ is computed by picking a point in $j$ and taking the four nearest points in $p$ as a basis for a quadratic interpolation. The normalisation factor $N_\text{over}$ is the total number of points used to evaluate $\mathcal{E}_p$. The results are summarised in Tab.~\ref{TablePBmini} and the corresponding curve collapse for Binder's cumulant is shown in Fig.~\ref{BinderRescaledAtLargeY}.

In principle, this method could also be used for $\chi_m$, but it would be necessary to minimise for five parameters. Our investigation shows that this leads to numerical instabilities and the extraction of critical exponents from the susceptibility using this method is not possible hitherto. \\
\begin{table}[H]
	\centering{
	\begin{tabular}{|c||c|c||c|}\hline
		& $T_c^{(L=\infty)}$ & $\nu$ & interval \\ \hline\hline
		$\kappa=0.06$ & 18.147(24) & 0.550(1) & 17.4, 18.8 \\ \hline   
		$\kappa=0.00$ & 16.667(27) & 0.525(6) & 16.0, 17.2 \\ \hline  
		O(4)                    & 0.3005(34) & 0.50000(3) & 0.294, 0.314 \\ \hline 
	\end{tabular}}
\caption[Curve collapse results of Binder cumulant.]{Curve collapse
  results of Binder's cumulant where the last column indicates the
  interval of the control parameter in which the procedure has been used. The parameter $T$ stands either for $y$ in the Higgs-Yukawa model or for $\kappa$ in the O(4) model. All errors are statistical only.}
\label{TablePBmini}
\end{table}

At this point we can claim that we have found a second order phase transition between 
the SYM and the FM phases in the strong Yukawa coupling regime. 
The absence of discontinuities in $\left<m\right>$ and the second-order finite size scaling of $\chi_m$ are strong evidence for such a statement. It is interesting to compare the critical exponents extracted from the susceptibility and Binder's cumulant with the ones of the weak-Yukawa model and the O(4) model. 

To be able to make a direct comparison of the O(4) model with the Higgs-Yukawa model, the same strategy 
has been used to compute observables in the pure scalar sector 
of the theory. In particular, the same analysis techniques have been used. 
The results of the correlated fit to $\chi_m$ are summarised in Tab.~\ref{TableFitSus}. 
The errors quoted there are purely statistical. Investigation of the dependence of the results 
on the fit interval leads to systematic uncertainties which are as large as the statistical errors roughly. It is not possible to claim a significant difference in the critical exponents between the Higgs-Yukawa model and the O(4) model from this method so far.

The curve collapse method, however, can only provide us with information 
about one critical exponent, namely $\nu$. The advantage of this method is 
the significantly smaller statistical error compared to the fit to $\chi_m$. 
However, it must be used with care. The scaling behaviour described in 
Eq.~(\ref{eq:Binder_cumulant_scaling}) is only true close to the 
critical point. If this method is applied at points too far away form 
the phase transition the result can be affected by scaling-violation effects. 
One possibility to achieve an impression of these effects is the  dependence on 
the interval in which the curve collapse method is applied. It was found, that 
the systematic uncertainty is roughly a factor of five larger than the 
statistical error. However, in the case of $\kappa=0.06$ and of the 
O(4) model the total error is still a factor of five smaller compared to the fitting procedure. In the case of $\kappa=0.00$ the total errors are compatible. 


The results of the critical exponent, $\nu$, in Tabs.~\ref{TableFitSus} and 
\ref{TablePBmini} indicates that the strong-Yukawa model and 
the O(4) model may belong to different universality classes.  
However, in the procedure of using Eq.~(\ref{ScalingFunction}) to 
determine this exponent, the difference of the two models can be as 
small as two standard deviations.  We stress that it is also important to 
investigate the scaling violation as pointed out in 
Refs.~\cite{Privman:1983dj,Binder:1985FSTH,Brezin:1981gm,Brezin:1985xx,Bernreuther:1987hv,Kenna:1992np,Kenna:2004cm}.
In particular, the observation of the multiplicative logarithmic scaling violation is 
directly related to the triviality of the 
theory~\cite{Aizenman:1983bz,Brezin:1981gm,Brezin:1985xx,Bernreuther:1987hv,Kenna:1992np,Kenna:2004cm}.  
Presently, we are exploring such analyses and performing computations at additional 
parameter values.  In the near future, we will therefore be able to see whether 
the value of $\nu$ in the strong bare Yukawa coupling regime is indeed 
different from the one of pure O(4) model.  
If we would find a significant difference, then it will be
important  
to investigate the strong-coupling regime closer and, in particular, 
a computation of the spectrum of the Higgs-Yukawa model in the strong-coupling region will 
become most interesting.

\subsection{Finite-temperature phase transition}\label{sec:finite_t}
One important subject
in the study of the Higgs-Yukawa model is  
the finite-temperature phase transition.  
In this section we describe the status of our investigation
of this transition. 
We are particularly interested in determining the critical temperature where the system undergoes a phase transition from the symmetric phase with vanishing 
scalar $vev$, $v=0$, to the broken phase with non-vanishing $v$. Further
interest lies in the determination of the order of the phase
transition and the critical exponents.
Preliminary results reported in this article are obtained at two values of the fermion mass, 
$m_f\approx 175\mathrm{GeV}$ and $m_f\approx 700\mathrm{GeV}$. 

Choosing the boundary conditions in the Euclidean temporal direction to be 
periodic for bosonic and anti-periodic for fermion fields, the 
temperature $T$ on the lattice is given by 
\begin{equation}
\label{eq:T_to_Lt}
 T = \frac{1}{a L_t} = \frac{\Lambda}{L_t}
\end{equation}
where $L_t$ denotes the dimensionless temporal extent of the lattice. 
For the study of the finite-temperature phase transition, 
we work at fixed bare Yukawa couplings which lead to the desired
fermion masses.  Results presented here are from lattice simulations
performed at $\hat{\lambda} = \infty$.  To vary the temperature, we
change the value of $\kappa$ at fixed $L_{t}$.  This is equivalent to
adjusting the lattice spacing while fixing the number of points in the
temporal extent of the lattice corresponding then to a change in 
the temperature.

Our study shows that the finite-temperature phase transitions in the 
Higgs-Yukawa model are consistent with second-order..  The order parameter is the
magnetisation as defined in Eqs.~(\ref{eq:bare_vev_definition}) and
(\ref{eq:mag}).  Since the correlation length is never divergent
because of finite-volume effects,  we resort to finite-size scaling
techniques to investigate the second-order finite-temperature phase
transition in this work.  In particular, we analyse the scaling
behaviour of the 
susceptibility of the magnetisation, Eq.~(\ref{eq:susc_def}).
As in Ref.~\cite{Jansen:1989gd}, we fit the 
susceptibility according to the partly phenomenologically motivated function
\begin{equation} \label{eq:naive_susceptibility_fit}
 \chi_m(\kappa) = A\left(L_s^{-2/\nu} + B_{_{+/-}}(\kappa - \kappa_c)^2\right)^{-\gamma/2}, \quad \nu=0.68,    \quad \gamma=1.38,
\end{equation}
where $A$, $B_{_{+/-}}$, and $\kappa_c$ are free fit parameters
($B_{_{+/-}}$ are coefficients in the broken and the symmetric phases, respectively), and 
$\nu$ and $\gamma$ are the 
critical exponents of the three dimensional 
O(4) model which are expected to characterise the 
second-order phase transition. 
Note that we use the fit function of Eq.~(\ref{eq:naive_susceptibility_fit}) with fixed values
of the critical exponents only to extract the critical value of
$\kappa$, denoted as $\kappa_c$ which in turn leads to the 
evaluation of the critical temperature. This approach is different from that
used for the investigation of the strong-Yukawa model as described in
Sec.~\ref{sec:bulk_phase_transition}.  Since 
$\kappa_c$ depends on the spatial volume, we perform simulations on 
various spatial lattice sizes and perform an infinite volume 
extrapolation using the formula ($D$ is an unknown constant),
\begin{equation} 
\label{eq:finite_volume_kappa_critical}
 \kappa_c(L) = \kappa_c(\infty) + D\cdot L^{-\nu}.
\end{equation}

Having extracted $\kappa_c$ in the infinite-volume limit,  $\kappa_{c}(\infty)$,
we can determine the lattice
spacing at this $\kappa$ value by  
performing zero-temperature simulations at exactly the same choice of
couplings, and using Eq.~(\ref{eq:setting_a}).   This then allows us
to predict the critical temperature, $T_{c}$,  through Eq.~(\ref{eq:T_to_Lt}).
In order to estimate the systematic effects in $T_{c}$ arising from
the uncertainty in $\kappa_{c}$, we also carry out two additional
zero-temperature simulations with $\kappa$ values chosen to reflect
the error on $\kappa_{c}$.
In this procedure, it is very challenging to maintain a constant Higgs
boson mass, since it depends significantly on the $\kappa$ value.
So far, we have not yet performed zero temperature runs for the presented results,
but from the results found in~\cite{Gerhold:2010bh} it is possible, to give a first estimate of the 
order of magnitude for the critical temperature and the corresponding Higgs boson masses in the case of a 
physical top quark mass.

\subsubsection{Finite-temperature study at physical top quark mass}
As the first step,  
we investigate the case of a degenerate fermion doublet with the quark mass close to 
the physical top quark mass.  To this end we 
fix the bare Yukawa coupling according to the tree-level 
estimate of $y = m_t/v_r$, which has been shown to be a 
good approximation in this region of couplings~\cite{Gerhold:2010bh}. 
We perform simulations at two different temporal extents ($L_t=4,6$)
for estimating the
discretisation effects.   In addition, three spatial extents,
$L_{s}=16,20,24$, are implemented in order to  
perform the infinite-volume extrapolation.

\begin{figure}[H]
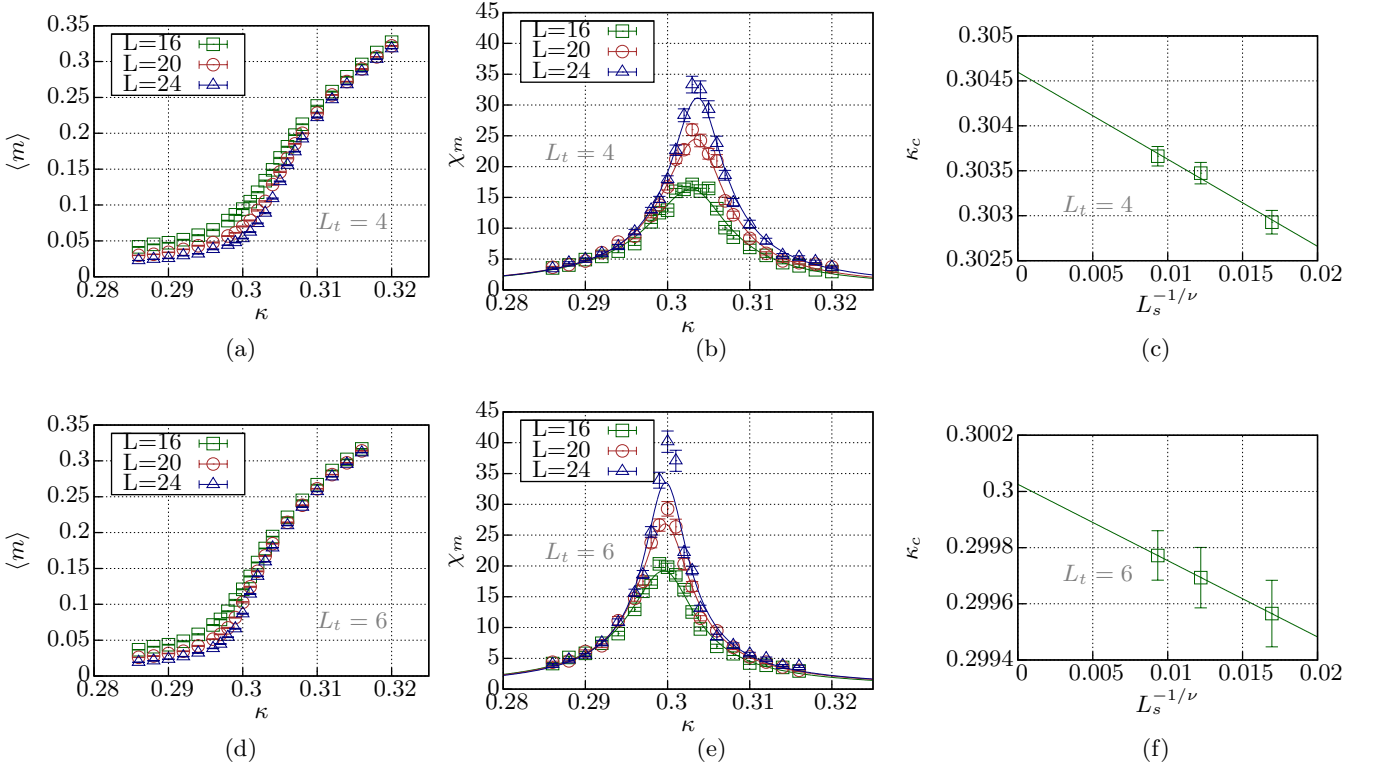

     \begin{center}
     $
     \begin{array}{ccc}
         \hspace*{-0.5cm} 
\input{figures/finiteT_2/finiteT_mt175_T4_mag.tex} &
         \hspace*{-1.2cm} 
\input{figures/finiteT_2/finiteT_mt175_T4_susc.tex} &
         \hspace*{-1.2cm} 
\input{figures/finiteT_2/finiteT_mt175_T4_kc_extra.tex} \vspace{-0.5cm} \\
  \text{(a)} & \text{(b)} & \text{(c)}  \\ 
          \hspace*{-0.5cm} 
\input{figures/finiteT_2/finiteT_mt175_T6_mag.tex} &
         \hspace*{-1.2cm} 
\input{figures/finiteT_2/finiteT_mt175_T6_susc.tex} &
         \hspace*{-1.2cm} 
\input{figures/finiteT_2/finiteT_mt175_T6_kc_extra.tex} \vspace{-0.5cm} \\
  \text{(d)} & \text{(e)} & \text{(f)}  \vspace {0.4cm}\\
     \end{array} 
     $
     \end{center}
\vspace*{-1.0cm}
\caption{Results of our finite-temperature study at the physical top quark
  mass with the quartic coupling $\hat{\lambda} = \infty$. 
Plots (a) and (d): The magnetisation for temporal extents of $L_t=4$ and $L_t=6$. 
Plots (b) and (e): The corresponding susceptibilities with the fit
function in Eq.~\eqref{eq:naive_susceptibility_fit}. 
Plot (c) and (f): Infinite-volume 
extrapolation of $\kappa_c$ using
Eq.~\eqref{eq:finite_volume_kappa_critical}. 
Note that for the case of zero temperature $L^2$ denotes 
$\sqrt{V_4}$ with $V_4=L_s^3 L_t$ and $L_t=2L_s$.} 
\label{fig:finiteT_mt175_summary}
\end{figure}

The results of the magnetisation at $L_{t} = 4$ and $6$ are plotted in 
Fig.~\ref{fig:finiteT_mt175_summary}(a) and
Fig.~\ref{fig:finiteT_mt175_summary}(d), respectively.   It is obvious that there is a transition 
from the symmetric to the broken phase for each choice of $L_{t}$.  The 
corresponding 
susceptibilities are shown in
Figs.~\ref{fig:finiteT_mt175_summary}(b) and 
\ref{fig:finiteT_mt175_summary}(e). 
The $L{-}$dependence of $\kappa_c(L)$ is well described by Eq.~(\ref{eq:finite_volume_kappa_critical}),
as can be seen in Fig.~\ref{fig:finiteT_mt175_summary}(c) and (f).  
%

Our finite-temperature study at a fermion mass close to 
physical top-quark mass is an on-going project at an early stage.   
Presently, the
simulations using $L_{t}=4$ and $6$ both result in the Higgs boson mass,
$m_{H} \sim 600$GeV, and the critical temperature, $T_{c} \sim 400$GeV. Those values are obtained from $\kappa_c$:
\begin{equation}\label{eq_kc_mt175}
 \kappa_c(\infty, L_t=4) = 30460(29)\quad\quad \kappa_c(\infty, L_t=6) = 0.30003(25)
\end{equation}
by a comparison with the results shown in~\cite{Gerhold:2010bh}.
To make our predictions more precise, we are performing additional
lattice computations.  In
particular, we are planning zero-temperature simulations with larger
spatial extent.   This will allow us to have better control of the
infinite-volume extrapolation.

%

\vspace{-0.0cm}

\subsubsection{Status of finite-temperature study at a quark mass of about 700GeV}
In this section we present the status of our work on the critical
temperature in the Higgs-Yukawa model with one heavy fermion doublet with a
mass of about 700GeV.
We follow the same strategy as in the previous section. 
Here
the zero-temperature simulations are still in progress.  Thus,
the lattice spacings for this calculation are not yet available to us.

\begin{figure}[H]
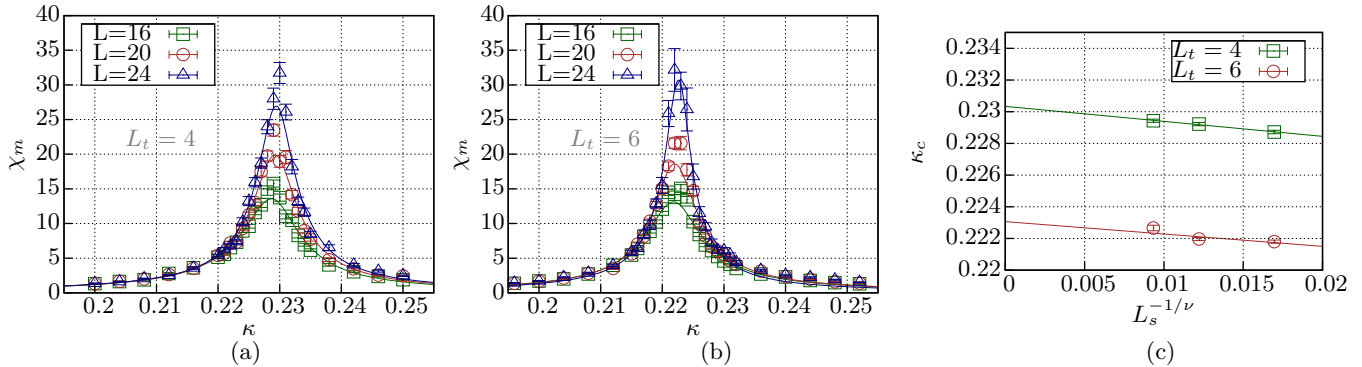

     \begin{center}
     $
     \begin{array}{ccc}
         \hspace*{-0.5cm} 
\input{figures/finiteT_2/finiteT_mt700_T4_susc.tex} &
         \hspace*{-1.2cm} 
\input{figures/finiteT_2/finiteT_mt700_T6_susc.tex} &
         \hspace*{-1.2cm} 
\input{figures/finiteT_2/finiteT_mt700_allT_kc_extra.tex} \vspace{-0.5cm} \\
  \text{(a)} & \text{(b)} & \text{(c)}  \vspace {0.4cm}\\
     \end{array} 
     $
     \end{center}
\vspace*{-1.0cm}
\caption{Plots (a) and (b) show the susceptibility as function of $\kappa$ at the 
large fermion mass of about $700\mathrm{GeV}$.  
Plot~(c) is the infinite-volume extrapolation for $\kappa_{c}$.}
\label{fig:finiteT_mt675_summary}
\end{figure}

Results of the susceptibility, and the infinite-volume extrapolation
for $\kappa_c$ can be found in Fig.~\ref{fig:finiteT_mt675_summary}.
From the phase structure presented in
Fig.~\ref{fig:phaseDiagramSummary} and the value of 
$\hat{y}\sim 2.8$, it is clear that the critical value 
of $\kappa$ is in the FM phase of the zero temperature theory, 
as expected.   We also notice that the values of $\kappa_{c}$ in the $L_{t}=6$ calculation
are smaller than that in the $L_{t}=4$ analysis.   This means that the $L_{t} = 6$
simulations are carried out closer to the FM${-}$SYM phase boundary,
and are thus performed at 
larger values of the cut-off.

%
%
%
%
\section{Outlook}\label{sec:summary}
In this article we have provided an overview
of non-perturbative lattice calculations of the Higgs-Yukawa sector
of the Standard Model and its extension with a fourth fermion 
generation. 
The phase diagram of the model has been studied and a complex  
and interesting structure has been revealed. At small values of the bare 
Yukawa coupling the properties of the phase transitions are 
consistent with the standard model expectation~\cite{Gerhold:2007yb,Gerhold:2007gx}.
However, we also establish an additional phase transition at very large values
of the bare Yukawa coupling~\cite{Gerhold:2007yb,Gerhold:2007gx,Bulava:2011jp}. 
This offers the very interesting possibility 
to investigate a strongly interacting Higgs-Yukawa model. 
We performed a detailed study of the properties of the phase transitions
at strong bare Yukawa coupling 
and determined the critical exponents characterising the phase transitions 
through a finite size scaling analysis. 
Although there are presently indications that these critical 
exponents may differ from the standard model ones, at this stage 
of our investigations 
it is too early to say that in the strong bare Yukawa coupling 
region indeed a non-Standard-Model-like phase structure exists.  

As an interesting direction we have also examined the Higgs-Yukawa model 
at non-zero temperature for fermion masses ranging from 
$175\text{GeV}$ to $700\text{GeV}$~\cite{Bulava:2011ss}. We find that the transition is always 
of second order and that the critical temperature is higher 
for increasing fermion mass. 

For a Standard Model top quark mass we have established lower and 
upper Higgs boson mass bounds as a function of the (lattice) cut-off 
of the theory~\cite{Gerhold:2009ub,Gerhold:2010bh,Gerhold:2010wy}. 
We also performed a detailed resonance analysis of the 
Higgs boson which confirmed that the Higgs boson mass bounds which 
assumed a stable Higgs boson are not affected by the resonance character 
of the Higgs boson~\cite{Gerhold:2011mx}. 
Furthermore, we find that the Higgs boson decay width 
into massive Goldstone bosons is never larger than 10\% of the Higgs boson 
mass and in good agreement with perturbative estimates. 
As a consequence of our lattice study of the lower and 
upper Higgs boson mass bounds within the Higgs-Yukawa sector at a
physical value of the top quark mass, 
we can, in principle, estimate the energy scale at which the standard model has to break 
down.  

We extended the study of the Higgs boson mass bound to a possible fourth
generation of quarks considering fermion masses up to $700\text{GeV}$~\cite{Gerhold:2010wv}. 
We found that the upper Higgs boson mass bound shows only a moderate shift by about 
20\% at such a fermion mass when compared to the bound for a 
Standard Model top quark mass.
However, the lower Higgs boson mass bound is altered significantly 
and can be as high as $500\text{GeV}$ for a fermion mass of $700\text{GeV}$. 
We complemented 
our non-perturbative lattice simulations with a lattice perturbative 
calculation of the lower Higgs boson mass bound from the effective 
potential. We found very good 
agreement with the lattice simulation data. This enabled us 
to test the stability of the lower bound against additions of 
higher dimensional operators. As a result we observed that the lower 
bound is not affected by including such additional operators. 
This finding puts severe constraints on the fourth generation 
if the particle with a mass of $125\text{GeV}$ seen at the LHC is 
the standard model Higgs boson. 

Let us discuss the consequences of 
our lattice study of the Higgs-Yukawa sector of the standard model 
and its extension to a fourth fermion generation, assuming that 
the particle detected at the LHC~\cite{ATLAS,CMS} is a 
Higgs boson with a mass of $125\text{GeV}$. For the standard model such a Higgs 
boson mass leads to rather small values of the renormalised 
quartic and Yukawa couplings and it seems therefore that the electroweak
sector of the standard model can be described perfectly 
within perturbation theory. Therefore, the perturbative analysis
of Ref.~\cite{Degrassi:2012ry} provides the result that the energy scale, up to which the 
Standard Model can be valid, is very high. Considering the extension of a fourth 
fermion generation, the lower Higgs boson mass bound together
with the phenomenological lower bound of the fourth generation fermion 
mass provides very severe constraints on the existence of the fourth
generation.

As a conclusion, our findings suggest that the electroweak theory of the Standard Model
is a perfect description of particle interaction up to very high energies as 
discussed in Ref.~\cite{Degrassi:2012ry}. 
Furthermore, a simple extension of the standard model by adding only a fourth 
fermion generation is most likely not realised. However, as discussed 
in Ref.~\cite{EliasMiro:2012ay} the addition of a singlet scalar field could change
the situation. As shown in Ref.~\cite{EliasMiro:2012ay}, the lower Higgs boson mass bound can be 
lowered significantly in the presence of such an  
additional scalar field. Of course, in Ref.~\cite{EliasMiro:2012ay} only a perturbative
calculation has been performed for the scenario of adding such a singlet 
scalar field and non-perturbative calculations, such as the ones
presented here, to scrutinise this picture
are highly desirable. 

We have demonstrated that with lattice field theory 
techniques generic strongly interacting Higgs-Yukawa theories can be 
studied in a controlled and accurate way. This became possible 
through a conceptual breakthrough of formulating chiral invariant 
theories on the lattice together with a much improved understanding 
of systematic effects such as finite size effects or determining 
resonance parameters. Since in addition the existing computing power of 
present super computers is clearly adequate to perform calculations 
of Higgs-Yukawa models, lattice computations can contribute to 
our understanding of Higgs-Yukawa models, in particular in the 
strongly interacting regime.

\section*{Acknowledgments}
This work is supported by Taiwanese
NSC via grants 100-2745-M-002-002-ASP (Academic Summit Grant),
99-2112-M-009-004-MY3, 101-2811-M-033-008, and 101-2911-I-002-509, and
by the DFG through the DFG-project Mu932/4-4, and the JSPS Grant-in-Aid for Scientific
Research (S) number 22224003. Simulations have been performed at the SGI system HLRN-II at
the HLRN supercomputing service Berlin-Hannover, the PAX cluster at DESY-Zeuthen, and HPC
facilities at National Chiao-Tung University and National Taiwan
University.  We thank the Galileo Galilei Institute for Theoretical 
Physics for hospitality and the INFN for the partial support during the
completion of this work.


\bibliographystyle{Bibliography/style.bst}
\bibliography{refs}

\end{document}